\documentclass[a4paper,fleqn]{cas-dc}
\usepackage[numbers]{natbib}
\usepackage{amsmath,amsfonts}
\usepackage[algo2e,linesnumbered,lined,boxed,commentsnumbered,ruled]{algorithm2e}
\usepackage{array}
\usepackage[caption=false,font=normalsize,labelfont=sf,textfont=sf]{subfig}
\usepackage{textcomp}
\usepackage{stfloats}
\usepackage{url}
\usepackage{verbatim}
\usepackage{graphicx}
\usepackage{makecell}
\usepackage{multirow}
\usepackage{booktabs}
\usepackage{color}
\usepackage{xcolor}

\def\tsc#1{\csdef{#1}{\textsc{\lowercase{#1}}\xspace}}
\tsc{WGM}
\tsc{QE}
\begin{document}
\let\WriteBookmarks\relax
\def\floatpagepagefraction{1}
\def\textpagefraction{.001}
\let\printorcid\relax % 可去掉页面下方的ORCID(s)

% \shorttitle{<short title of the paper for running head>} 
\shorttitle{Analytic Personalized Federated Meta-Learning} 

% \shortauthors{<short author list for running head>}
\shortauthors{Shunxian Gu et al.}

%论文标题
\title[mode = title]{Analytic Personalized Federated Meta-Learning}  

%作者信息
\author[1,2]{Shunxian Gu}
\author[2]{Chaoqun You}
\author[1]{Deke Guo}
\cormark[1]
\author[3]{Zhihao Qu}
\author[1]{Bangbang Ren}
\author[3]{Zaipeng Xie}
\author[1]{Lailong Luo}

\address[1]{National University of Defense Technology, Changsha 410073, China} %声明第一单位
\address[2]{Fudan University, Shanghai 200438, China} %声明第二单位
\address[3]{Hohai University, Nanjing 211100, China}
\cortext[1]{Corresponding author}  %声明通讯作者

\begin{abstract}
Analytic Federated Learning (AFL) is an enhanced gradient-free federated learning (FL) paradigm designed to accelerate training by updating the global model in a single step with closed-form least-square (LS) solutions. However, the obtained global model suffers performance degradation across clients with heterogeneous data distribution. Meta-learning is a common approach to tackle this problem by delivering personalized local models for individual clients. Yet, integrating meta-learning with AFL presents significant challenges: First, conventional AFL frameworks cannot support deep neural network (DNN) training which can influence the fast adaption capability of meta-learning for complex FL tasks. Second, the existing meta-learning method requires gradient information, which is not involved in AFL. To overcome the first challenge, we propose an AFL framework, namely FedACnnL, in which a layer-wise DNN collaborative training method is designed by modeling the training of each layer as a distributed LS problem. For the second challenge, we further propose an analytic personalized federated meta-learning framework, namely pFedACnnL. 
It generates a personalized model for each client by analytically solving a local objective which bridges the gap between the global model and the individual data distribution.
FedACnnL is theoretically proven to require significantly shorter training time than the conventional FL frameworks on DNN training while the reduction ratio is $83\%\sim99\%$ in the experiment. Meanwhile, pFedACnnL excels at test accuracy with the vanilla FedACnnL by $4\%\sim8\%$ and it achieves state-of-the-art (SOTA) model performance in most cases of convex and non-convex settings compared with previous SOTA frameworks.
\end{abstract}


% \begin{highlights}
% \item highlight-1
% \item highlight-2
% \item highlight-3
% \end{highlights}

% Keywords

\maketitle

% Main text
\section{Introduction}
The rapid expansion of internet-of-things (IoT) devices, increasing from under 9 billion in 2019 to over 15 billion by 2023 \cite{ficco2024federated}, has been a key driver in the development of Federated Learning (FL). FL is a privacy-preserving technique of Distributed Collaborative Machine Learning (DCML) that enables model training across multiple edge devices without uploading raw data to a central parameter server (PS). Training a general FL model contains three typical steps: (i) a set of edge devices conduct local computation for model training, usually by the stochastic gradient descent (SGD) method, based on their own dataset, and upload the resultant model parameters to the server, (ii) the server aggregates the uploaded parameters to generate a new global model, and (iii) the server feeds back the new global model to edge devices for the next round of local model training. This procedure repeats until the loss function converges and a certain model training accuracy is achieved.

Typical FL training schemes include first- \cite{mcmahan_communication-efcient_nodate, liu2024fedcd} (e.g. SGD) and second-order methods \cite{sen2024fagh}, which necessitate the computation of gradients and Hessian matrices. Despite the efficiency of these gradient descent (GD)-based methods, their reliance on gradient computations can pose challenges in practice. Specifically, (i) edge devices with poor computational capabilities are incapable of performing frequent gradient calculations. For instance, mobile phones or IoT sensors are primarily designed for inference-only purposes rather than backpropagation \cite{feng2025baffle}. In such cases, only small, pre-trained or pruned models are computationally affordable for execution on these devices~\cite{ficco2024federated, wang2019haq}. (ii) Gradient information is not always available to support gradient computations. For instance, in situations such as federated hyperparameter tuning \cite{dai2020federated} and federated black-box attack \cite{yi2022zeroth}, the gradients cannot be derived due to the non-differentiable nature of the loss function. As a result, GD-based FL frameworks may become infeasible in these scenarios, prompting a strong demand from both academia and industry for the development of gradient-free FL training approaches.

Recently, the zeroth-order FL framework has been widely adopted as an efficient gradient-free FL training approach \cite{feng2025baffle,fang2022communication, chen2024fine}. The conventional zeroth-order FL frameworks, including BAFFLE~\cite{feng2025baffle} and FedZO~\cite{fang2022communication}, conduct the finite difference technique to approximate gradients, effectively substituting the backpropagation in gradient computations with multiple inference processes. 
However, such gradient approximation is always biased, necessitating intensive sampling processes to reduce the approximation error. Consequently, the training time required by these frameworks is excessively long. To solve this problem, analytic federated learning (AFL) \cite{zhuang2024analytic} has been proposed as an effective and emerging approach without any gradient approximation or computations. 

In AFL, the global model is updated only once by using closed-form least-square (LS) solutions. Therefore, AFL can achieve a significantly shorter training time than the conventional zeroth-order FL frameworks. However, it overlooks the performance limitation of training a single global model for clients with heterogeneous data distributions. In other words, training a unified global model to satisfy the performance requirements of all clients is challenging due to the distinct preferences in data generation or collection by different clients. Personalized FL (PFL) \cite{sabah2024model} is a promising paradigm to alleviate the heterogeneous data distribution problem. In PFL, each client utilizes the common knowledge, which is the aggregated global model on the PS, to generate a personalized model tailored for its local data distribution. Meta-learning \cite{gao2023configure, yang2023personalized, t2020personalized}, as a representative model optimization technique in PFL, enables the aggregated global model (i.e. the meta-model) to fast adapt to heterogeneous tasks by performing several gradient descents on the global model. This motivates us to introduce meta-learning to AFL to address the heterogeneous data distribution problem in AFL. This introduction results in two challenges: First of all, the current AFL framework \cite{zhuang2024analytic} does not have the capability of training deep neural networks (DNN). This drawback can influence the fast adaption capability of meta-learning for complex machine learning tasks. Secondly, the existing meta-learning method is only suitable for clients using GD-based optimization methods. Therefore, combining meta-learning with AFL directly is impractical since AFL does not involve any gradient approximation or computations.

In this article, we first resort to ACnnL (analytic convolutional neural network learning) \cite{zhuang2025analytic}, a single-device analytic learning method for DNN, and propose its federated implementation, called FedACnnL, to enable DNN training in AFL. Specifically, we model the training of each layer as a distributed LS problem and design a layer-wise global model update rule using closed-form LS solutions to achieve a layer-wise DNN training process identical to that of ACnnL, under the federated setting. Inheriting from FedACnnL, we propose an analytic personalized federated meta-learning (APFML) framework, called pFedACnnL. A gradient-free meta-learning method is designed in it to combat the heterogeneous data distribution problem. Specifically, the PS groups the clients with similar data distribution and a meta-model that contains common knowledge is generated within each group by following the standard training process of FedACnnL. Next, each client can achieve its personalized model tailored for its local data distribution without using any gradient information by analytically solving a local PFL objective. This objective bridges the gap between the personal knowledge of its local dataset and the common knowledge of its meta-model. Our contributions can be summarized as follows:

\begin{itemize}
    \item We first resort to ACnnL and propose FedACnnL to enable DNN training in AFL. With a competitive test accuracy of the trained DNN model, FedACnnL is proven to require a greatly shorter training time than the conventional zeroth-order FL frameworks both theoretically and experimentally.
    \item Based on FedACnnL, we propose pFedACnnL, which is the first APFML framework to the best of our knowledge. A gradient-free meta-learning method is designed in it for fast adaption of the group-wise meta-models to local tasks, thereby promoting the model performance on each client under the heterogeneous data distribution setting.
    \item We validate the efficiency of our frameworks on a CPU-cluster-based testbed, which can emulate the environmental settings of data heterogeneity and resource heterogeneity in FL simultaneously. Experimentally, FedACnnL can reduce the training time by $83\%\sim99\%$ compared with the conventional first- and zeroth-order FL frameworks. Meanwhile, pFedACnnL achieves state-of-the-art (SOTA) model performance in most cases of convex and non-convex settings, compared with other previous SOTA first- and zeroth-order PFL frameworks.
\end{itemize}

\section{Preliminaries}\label{p2}
\subsection{Notations}
For convenience, we unify the notations used in this article as follows. Non-bold English and Greek letters (e.g. $N,i,,\gamma,\epsilon$) denote scalars. Bold lowercase English letters (e.g. $\textbf{x},\textbf{h}$) represent vectors while bold uppercase English letters (e.g. $\textbf{X},\textbf{Y}$) represent matrices. Letters in the calligraphic font (e.g. $\mathcal{S},\mathcal{Q},\mathcal{W}$) are used to represent sets. Finally, we define square bracket $[;\cdots;]$ as the concatenation operator, which concatenates multiple vectors into one. For instance, let $\textbf{A}_1\in\mathbb{R}^{a_1\times{b}}$, $\textbf{A}_2\in\mathbb{R}^{a_2\times{b}}$ and $\textbf{A}_3\in\mathbb{R}^{a_3\times{b}}$, we can have $\textbf{C}=[\textbf{A}_1;\textbf{A}_2;\textbf{A}_3]\in\mathbb{R}^{(a_1+a_2+a_3)\times{b}}$. Some notations that may cause confusion for readers are enumerated in the Table \ref{tab1}.

\begin{table*}[!tbp]\rmfamily
\renewcommand{\arraystretch}{1.27}
\setlength\tabcolsep{2.8pt}
\caption{Table of partial notations.}
\begin{center}
\centering
\begin{tabular}{|l|l|}
%添加顶部横线 
\Xhline{1.5 pt}
%输入标题
% \cite{2-3}
% \cline{2-7}
% \Xhline{1.5 pt}
%添加标题和内容之间的横线
% \Xhline{0.5 pt}
\centering
% \cline{2-4}
Notations&Description\\
\Xhline{0.5 pt}
$n$& the index of a sample\\
$i$& the index of a client\\
$b$& the index of a local batch\\
$\textbf{x}_n$&the $n$-th data vector of the entire dataset $\mathcal{S}$\\
$\textbf{X}$&the data matrix of the entire dataset $\mathcal{S}$\\
$\textbf{X}^{(i)}$&the data matrix of the local dataset on the $i$-th client\\
$\textbf{X}_l^{[n]}$&the real input matrix of the $l$-th layer using $\textbf{x}_n$ as DNN input\\
\multirow{2}{*}{$\textbf{X}_l^{(i)}$}&the real input matrix of the $l$-th layer \\&on the $i$-th client using $\textbf{X}^{(i)}$ as DNN input\\
$\textbf{X}^{\{b\}}$&the $b$-th batch of data matrix\\
$\textbf{X}_l^{\{b\}}$&the real input matrix of the $l$-th layer using $\textbf{X}^{\{b\}}$ as DNN input\\
$\textbf{y}_n$&the $n$-th label vector of the entire dataset $\mathcal{S}$\\
$\textbf{Y}$&the label matrix of the entire dataset $\mathcal{S}$\\
$\textbf{Y}^{(i)}$&the label matrix of the local dataset on the $i$-th client\\
$\bar{\textbf{Z}}_l^{[n]}$& the pseudo label matrix of the $l$-th layer by encoding $\textbf{y}_n$\\
\multirow{2}{*}{$\bar{\textbf{Z}}_l^{(i)}$}&the pseudo label matrix of the $l$-th layer \\&on the $i$-th client by encoding $\textbf{Y}^{(i)}$\\
$\textbf{Y}^{\{b\}}$&the $b$-th batch of label matrix\\
$\bar{\textbf{Z}}_l^{\{b\}}$&the pseudo label matrix of the $l$-th layer by encoding $\textbf{Y}^{\{b\}}$\\
$(\cdot)^T$& the transpose of a matrix\\
$\lceil\cdot\rceil$&the smallest integer greater than a given number\\
% $\textbf{X}_l^{\{i\}}$the real input matrix of the $l$-th layer\\
\Xhline{1.5 pt}
\end{tabular}
\label{tab1}
\end{center}
\vspace{-0.6cm}
\end{table*}

\subsection{Analytic Convolutional Neural Network Learning (ACnnL)}\label{p2p2}
Unlike the prior analytic learning methods \cite{toh2018learning, ranganathan2020zorb, lee2023gpil} that only support shallow DNN training or training the single output layer with a pre-trained backbone, ACnnL enables DNN training. AcnnL projects the label information into the hidden layers so that each hidden layer is equipped with supervised information (i.e. pseudo labels). Then, it updates the model weights layer-wisely by using closed-form LS solutions in a supervised learning style with the layer input and pseudo labels. Since the real label lies within the pseudo one’s linear space, mapping the input features onto the pseudo labels is almost equivalent to mapping it onto the real label, thereby achieving DNN training \cite{zhuang2025analytic}. Meanwhile, thanks to the layer-wise weight update mechanism, ACnnL completes model training in only one epoch, in which the number of iterations matches that of the DNN layers. In the following, we recap the training process in each iteration.

\begin{figure}[tbp]
\centering
\includegraphics[width=8.0cm]{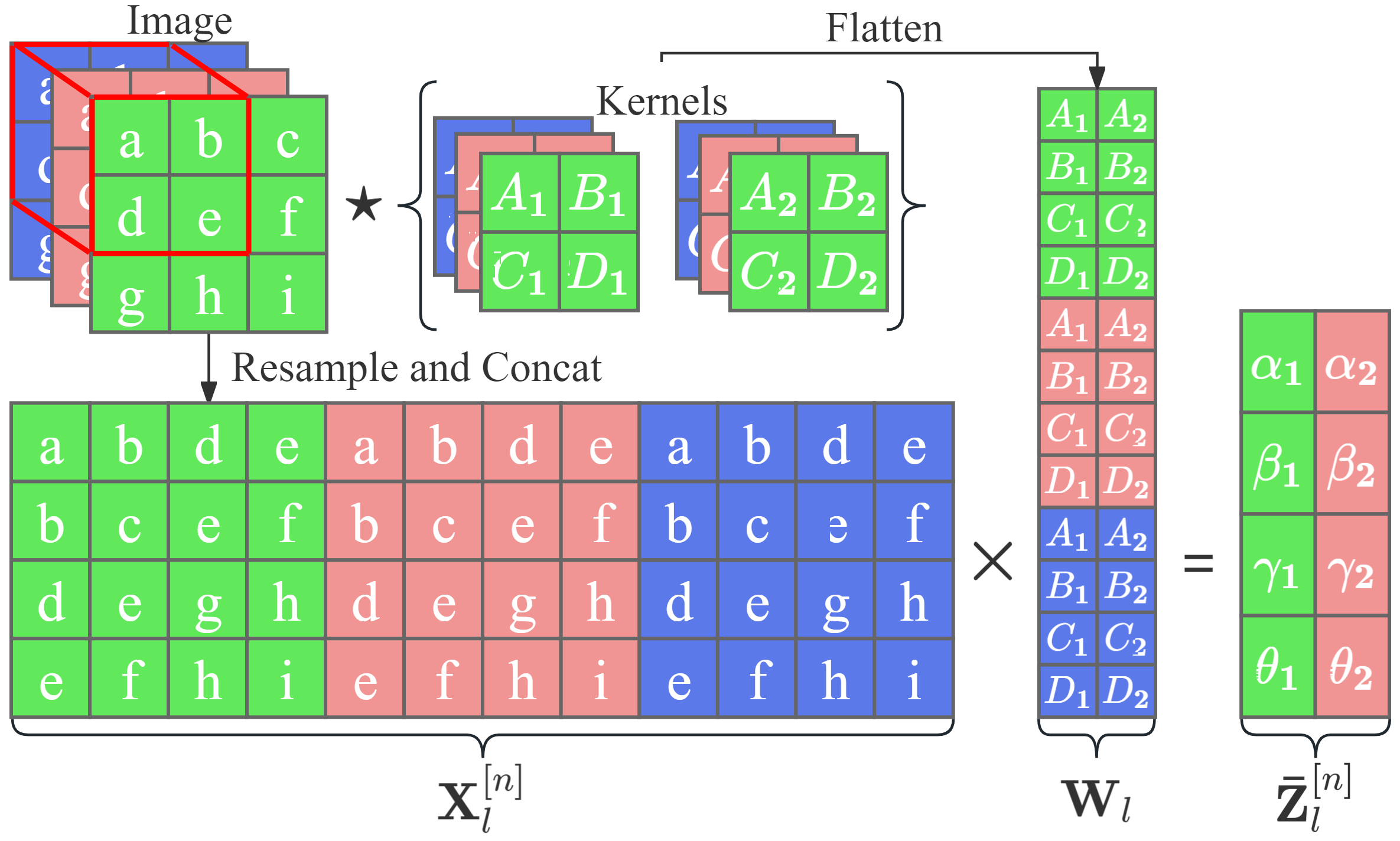}%
\caption{The resampling and flattening process in convolutional layers to make the weights in it updatable by the equation \ref{eq2}.}
\label{fig1}
\end{figure}

Given an $N$-sample dataset $\mathcal{S}=\{\textbf{x}_n,\textbf{y}_n\}_{n=1}^N$ where $\textbf{x}_n\in{\mathbb{R}}^{d_0}$ and $\textbf{y}_n\in{\mathbb{R}}^{d_y}$ denote the data vector and the label vector of the $i$-th sample respectively. We stack these samples to generate the data matrix $\textbf{X}=\left[\textbf{x}_1; \textbf{x}_2;\cdots;\textbf{x}_N\right]\in\mathbb{R}^{N\times{d_0}}$ and the label matrix $\textbf{Y}=\left[\textbf{y}_1;\textbf{y}_2;\cdots;\textbf{y}_N\right]\in\mathbb{R}^{N\times{d_y}}$. In the $l$-th iteration, the pseudo label matrix of the $l$-th hidden layer $\bar{\textbf{Z}}_l$ is generated as $\bar{\textbf{Z}}_l=\textbf{YQ}_l$, where ${\textbf{Q}_l}\in{\mathbb{R}^{d_y\times{d_l}}}$ is the label encoding matrix which is randomly initialized from a normal distribution. The encoding is linear so that $\bar{\textbf{Z}}_l$ can fit the dimensionality of $\textbf{Z}_l\in\mathbb{R}^{N\times{d_l}}$, which is the real output matrix before entering the activation function of the $l$-th hidden layer $f_l(\cdot)$ when $\textbf{X}$ is the input of the DNN model being trained. Specially, $\bar{\textbf{Z}}_l=\textbf{Y}$ when the output layer of the model is being trained.
Then, ACnnL formulates the supervised learning of the $l$-th layer as an least-square (LS) problem with $l_2$ regularization:
\begin{equation}
\underset{\textbf{W}_l}{\arg\min}~\Vert{\bar{\textbf{Z}}_l-\textbf{X}_l\textbf{W}_l}\Vert_2^2+\gamma\Vert{\textbf{W}_l}\Vert_2^2
\label{eq1}
\end{equation}
where $\textbf{X}_l\in\mathbb{R}^{N\times{d_{l-1}}}$ and $\textbf{W}_l\in\mathbb{R}^{d_{l-1}\times{d_l}}$ denote the real input matrix and model weights of the $l$-th hidden layer respectively. Specially, $\textbf{X}_l=\textbf{X}$ when the input layer is being trained. $\gamma$ is the regularization coefficient which is a hyperparameter to control the search space of the LS solutions. Next, we can take the closed-from LS solutions of the weights in $l$-th hidden layer by putting the derivatives with respect to $\textbf{W}_l$ on the objective function in the equation \ref{eq1} to 0. We present the solutions at a sample level as follows:
\begin{equation}
\hat{\textbf{W}}_l = (\sum_{n=1}^N{\textbf{X}_l^{[n]T}\textbf{X}_l^{[n]}}+\gamma{\textbf{I}})^{-1}(\sum_{n=1}^{N}\textbf{X}_l^{[n]T}\bar{\textbf{Z}}_l^{[n]})
\label{eq2}
\end{equation}
where $\hat{\textbf{W}}_l$ denotes the trained model weights of the $l$-th layer. Specially, when ACnnL trains convolutional layers, $\textbf{X}_l^{[n]}\in\mathbb{R}^{W_{l-1}\times{H_{l-1}}\times{C_{l-1}}}$ is an image and $\textbf{W}_l~(i.e.~{\mathcal{W}_l})=\{\textbf{W}_l^{[c]}\}_{c=1}^{C_l}, \textbf{W}_l^{[c]}\in\mathbb{R}^{ks\times{ks}\times{C_{l-1}}}$ is the set of convolution kernels, where $W,H,C$ are width, height and channel respectively, and $ks$ is the kernel size. As illustrated in Fig. \ref{fig1}, ACnnL reformulates the supervised learning of convolutional layers to be compatible with the equation \ref{eq1} by resampling on the $\textbf{X}_l^{[n]}$ with a sliding window of $ks$ size and flattening the $\mathcal{W}_l$ to $\textbf{W}_l\in\mathbb{R}^{(ks\times{ks}\times{C_{l-1}})\times{C_l}}$. After the flattened $\textbf{W}_l$ is replaced by $\hat{\textbf{W}}_l$ generated from the equation \ref{eq2}, $\hat{\textbf{W}}_l$ is transformed into $\hat{\mathcal{W}}_l$, which has the original shape of $\mathcal{W}_l$.

\begin{figure}[tbp]
\centering
\includegraphics[width=8.0cm]{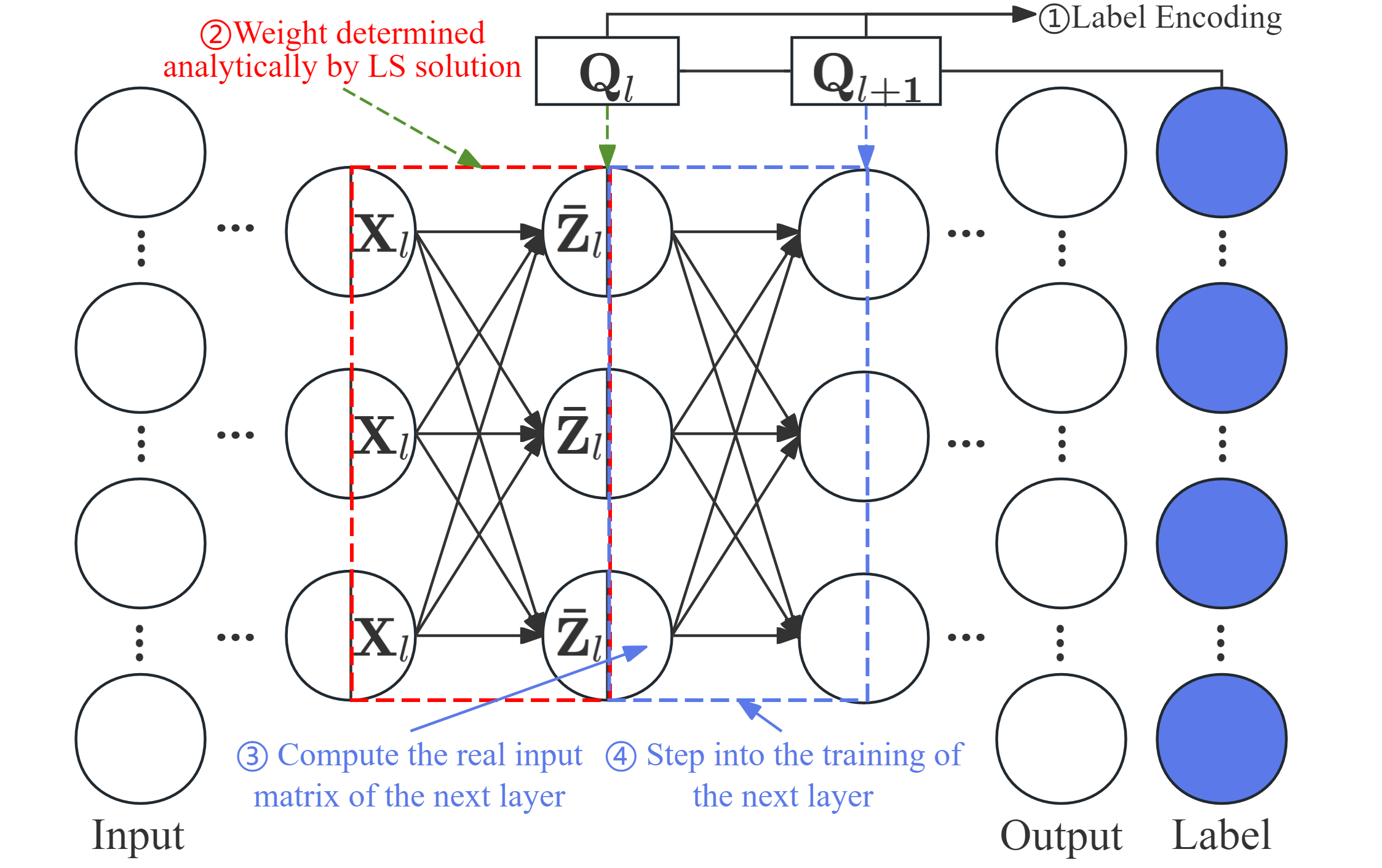}%
\caption{The workflow of ACnnL in the arbitrary iteration. The red dashed box represents the current training layer while the blue dashed box indicates the next layer to be trained.}
\label{fig2}
\end{figure}

After ACnnL completes the aforementioned label encoding stage and the weight update stage, it computes the real input matrix of $(l+1)$-th layer, $\textbf{X}_{l+1}=f_l(\textbf{X}_l\hat{\textbf{W}}_l)~or~f_l(\textbf{X}_l{\star}\hat{\mathcal{W}}_l)$ (where $\star$ is the convolution operation) and finally steps into the next iteration to train the next layer. ACnnL finishes its one-epoch model training when all layers of DNN are trained. Fig. \ref{fig2} pictures the workflow in the arbitrary iteration of ACnnL.

\section{Design of FedACnnL}\label{p3}
In this section, we first model the weight update problem in each layer of ACnnL as a distributed LS problem. Then, based on the distributed LS problem, we propose the federated implementation of ACnnL, called FedACnnL, followed by the adaptive batch size algorithm incorporated in it. Finally, we give a theoretical analysis on the complexity of the computation and communication in FedACnnL.
\subsection{Distributed LS Problem}\label{p2p3}
Based on the supervised learning property of ACnnL in each hidden layer, we can model the weight update problem in each layer as a distributed LS problem. Specifically, suppose we have the aforementioned $N$-sample dataset $\mathcal{S}$ in the section \ref{p2p2}, and $C$ clients of edge devices. The dataset is disjointly distributed onto the $C$ clients, i.e. $\mathcal{S}=\cup_{i=1}^C\mathcal{D}_i,\ {\ \mathcal{D}}_{m}\cap \mathcal{D}_{n}=\emptyset,\ \forall\ m\neq n$, where $i$ is the index of a client and $\mathcal{D}_i$ denotes the local dataset of a client. Each client stacks the local dataset to generate the local data matrix $\textbf{X}^{(i)}$ and the local label matrix $\textbf{Y}^{(i)}$ like section \ref{p2p2}. Then, in the arbitrary $l$-th hidden layer, each client encodes $\textbf{Y}^{(i)}$ with a same global $\textbf{Q}_l$ to generate a local pseudo label matrix $\bar{\textbf{Z}}_l^{(i)}=\textbf{Y}^{(i)}\textbf{Q}_l$. Specially, $\bar{\textbf{Z}}_l^{(i)}=\textbf{Y}^{(i)}$ when the $l$-th hidden layer is the output layer. With the real input matrix at the $l$-th layer $\textbf{X}_l^{(i)}$ and the pseudo label matrix $\bar{\textbf{Z}}_l^{(i)}$, we can formulate the distributed/federated optimization problem at the arbitrary $l$-th layer as:
\begin{equation}
    \begin{split}
    &\underset{\textbf{W}_l}{\arg\min}~\sum_{i=1}^{C}F(\textbf{W}_l;\mathcal{D}_i) \\
    where~F(\textbf{W}_l&;\mathcal{D}_i)=\Vert{\bar{\textbf{Z}}_l^{(i)}-\textbf{X}_l^{(i)}\textbf{W}_l}\Vert_2^2+\gamma\Vert{\textbf{W}_l}\Vert_2^2  \\
    \end{split}
\label{eq3}
\end{equation}
Assume that the model weights prior to the $l$-th layer on all clients are identical, in other words, all $\textbf{X}^{(i)}$ undergo the same computational operations to obtain $\textbf{X}_l^{(i)}$. Then, we can get the following two facts:
\begin{equation}
    \begin{split}
    &\sum_{i=1}^C{\textbf{X}_l^{(i)T}\textbf{X}_l^{(i)}}=\sum_{n=1}^N{\textbf{X}_l^{[n]T}\textbf{X}_l^{[n]}} \\
    &\sum_{i=1}^C\textbf{X}_l^{(i)T}\bar{\textbf{Z}}_l^{(i)}=\sum_{n=1}^{N}\textbf{X}_l^{[n]T}\bar{\textbf{Z}}_l^{[n]}\\
    \end{split}
\label{eq4}
\end{equation}
That is because the overall dataset formed by all local datasets of the clients is equivalent to the original $N$-sample dataset and the sum of all intermediate output matrices is not changed. Finally, the federated optimization problem can be analytically solved using the equation \ref{eq2} and the equation \ref{eq4}:
\begin{equation}
\hat{\textbf{W}}_l = (\underbrace{\sum_{i=1}^C{\textbf{X}_l^{(i)T}\textbf{X}_l^{(i)}}}_{\textbf{A}_0}+\gamma{\textbf{I}})^{-1}(\underbrace{\sum_{i=1}^{C}\textbf{X}_l^{(i)T}\bar{\textbf{Z}}_l^{(i)}}_{\textbf{A}_1})
\label{eq5}
\end{equation}
For convenience, we define the sum results of the two facts as $\textbf{A}_0$ and $\textbf{A}_1$ respectively in the rest of the article. Based on the distributed LS problem, we propose the federated implementation of ACnnL, i.e. FedACnnL, in the next subsection.
\subsection{Federated Implementation of ACnnL}\label{p3p1}
Before we step into the description of the training rounds with a same workflow, we first present the \textbf{initialization} at the beginning of FedACnnL: Firstly, each client initializes a local DNN model with the same structure. Then, the PS broadcasts the same random seed $s$ to all clients. Finally, each client utilizes $s$ to generate an identical set of label encoding matrices, i.e. $\mathcal{Q}=\{\textbf{Q}_l\}_{l=1}^{L-1}$, where $L$ is the total number of the DNN layers being trained.

Based on the distributed LS problem defined above, we give a detailed description of the workflow in the arbitrary $l$-th training round, in which clients collaboratively train the $l$-th hidden layer of the DNN model, as follows:

\textbf{Local Computation.} Firstly, each client allocates the memory space for $\textbf{X}_l^{(i)T}\textbf{X}_l^{(i)}$ and $\textbf{X}_l^{(i)T}\bar{\textbf{Z}}_l^{(i)}$, by creating zero matrices with the same shapes of them. Then, each client randomly samples a fixed batch of data matrix i.e. $\textbf{X}^{\{b\}}\in\mathbb{R}^{B_l^{(i)}\times{d_0}}$ and label matrix $\textbf{Y}^{\{b\}}\in\mathbb{R}^{B_l^{(i)}\times{d_y}}$, where $b$ is the index of the local batch and $B_l^{(i)}$ is the local batch size of the $i$-th client in the $l$-th training round, from the local dataset $\mathcal{D}_i$ without replacement until no batch of data can be generated from it. In each sampling process, each client utilizes the previously trained layers, i.e. $\{\hat{\textbf{W}}_1,\cdots,\hat{\textbf{W}}_{l-1}\}$, which are identical over all clients, to generate a batch of real input matrix $\textbf{X}_l^{\{b\}}$ by the forward pass of $\textbf{X}^{\{b\}}$ from the input layer to the $(l-1)$-th layer. Meanwhile, the pseudo label matrix $\bar{\textbf{Z}}_l^{\{b\}}$ is obtained by the label encoding technique, i.e. $\bar{\textbf{Z}}_l^{\{b\}}=\textbf{Y}^{\{b\}}\textbf{Q}_l$ (or $\bar{\textbf{Z}}_l^{\{b\}}=\textbf{Y}^{\{b\}}$ when the $l$-th layer is the output layer). The $\textbf{X}_l^{\{b\}T}\textbf{X}_l^{\{b\}}$ and $\textbf{X}_l^{\{b\}T}\bar{\textbf{Z}}_l^{\{b\}}$ is added to the $\textbf{X}_l^{(i)T}\textbf{X}_l^{(i)}$ and $\textbf{X}_l^{(i)T}\bar{\textbf{Z}}_l^{(i)}$ respectively at the end of each sampling process. To summarize, the local computation process of the $l$-th training round can be mathematically represented as:
\begin{equation}
    \begin{split}
    &\textbf{X}_l^{(i)T}\textbf{X}_l^{(i)}=\sum_{b=1}^{\lceil|\mathcal{D}_i|/B_l^{(i)}\rceil}{\textbf{X}_l^{\{b\}T}\textbf{X}_l^{\{b\}}} \\
&\textbf{X}_l^{(i)T}\bar{\textbf{Z}}_l^{(i)}=\sum_{b=1}^{\lceil|\mathcal{D}_i|/B_l^{(i)}\rceil}\textbf{X}_l^{\{b\}T}\bar{\textbf{Z}}_l^{\{b\}}\\
    \end{split}
\label{eq6}
\end{equation}
where $|\mathcal{D}_i|$ is the total number of samples in the local dataset of the $i$-th client. There exists a fact in the equation \ref{eq6} that regardless of what number $B_l^{(i)}$ is, the resultant $\textbf{X}_l^{(i)T}\textbf{X}_l^{(i)}$ and $\textbf{X}_l^{(i)T}\bar{\textbf{Z}}_l^{(i)}$ remain invariant. This fact indicates the \textit{weight-invariant property} of AFL, which is further discussed in the latter of this subsection.

\textbf{Aggregation.} The resultant $\textbf{X}_l^{(i)T}\textbf{X}_l^{(i)}$ and $\textbf{X}_l^{(i)T}\bar{\textbf{Z}}_l^{(i)}$ are uploaded to the PS by each client. After the PS receives the resultant two items from all clients, it aggregates the items respectively by a sum operation, i.e. $\sum_{i=1}^C{\textbf{X}_l^{(i)T}\textbf{X}_l^{(i)}}$ and $\sum_{i=1}^{C}\textbf{X}_l^{(i)T}\bar{\textbf{Z}}_l^{(i)}$, to generate the $\textbf{A}_0$ and $\textbf{A}_1$ in the equation \ref{eq5}. Then, the PS utilizes the $\textbf{A}_0$ and $\textbf{A}_1$ to update the global model weights of the $l$-th layer by using the closed-from LS solutions in the equation \ref{eq5}. Finally, the PS feeds the trained global model weights at the $l$-th layer, i.e. $\hat{\textbf{W}}_l$, back to all clients.

\textbf{Local Weight Update.} After clients receive the $\hat{\textbf{W}}_l$ from the PS, they replace the local model weights of the $l$-th layer with the global one, i.e. $\hat{\textbf{W}}_l$. Then, they step into the next training round to collaboratively train the next layer, i.e. $(l+1)$-th layer until all layers of their local DNN model have been trained.

\textbf{Weight-Invariant Property.} Based on the facts in the equation \ref{eq4} and \ref{eq6}, we can find that the $\textbf{A}_0$ and $\textbf{A}_1$ in the equation \ref{eq5} remain invariant unless the overall dataset formed by all local datasets of the clients, or the set of label encoding matrices $\mathcal{Q}$ changes. Therefore, no matter how the overall dataset is distributed among the clients, no matter what number the batch size is, and no matter how the model weights are initialized on the clients, the final closed-form LS solutions of each layer will not be changed in FedACnnL. This property is utilized in the following adaptive batch size algorithm.

\subsection{Adaptive Batch Size Algorithm}
In the real-world FL scenario, the computing power and the network conditions of the clients are various. Thus, the overall time consumption of the clients completing the local computation and uploading the aforementioned resultant two items are inconsistent. This inconsistency is the straggler phenomenon of FL, in which the PS must keep idle until the resultant items of the client with the poorest computing power and network condition arrive at the PS. An empirical practice to alleviate this phenomenon is to increase the local batch size of the lagging clients to fully exert the parallelism of the hardware (CPUs or GPUs), thereby reducing the local computation time of the lagging clients. As a result, the item arrival time of the lagging clients at the PS can be advanced, reducing the idle duration of the PS. In our design, the PS adaptively adjusts the local batch size for each client in each training round by computing the ratio of each client's current time consumption to the global minimum one of the first round. The adjusted result is then sent back to each client with the trained global model weights obtained in the aggregation step. This adjustment can be mathematically represented as:
\begin{equation}
B_{l+1}^{(i)}=min(\lceil\frac{T_l^{(i)}-T_{l-1}^{(i)}}{min(\{T_1^{(i)}-T_{0}^{(i)}\}_{i=1}^C)}\times{B_{1}^{(i)}}\rceil,B_{max}^{(i)})
\label{eq7}
\end{equation}
where $T_l^{(i)}$ denotes the item arrival time of the $i$-th client at the PS in the $l$-th training round and $T_0^{(i)}=0$. The time interval between two consecutive item arrivals of the $i$-th client, i.e. $T_l^{(i)}-T_{l-1}^{(i)}$ is an approximation of the client's overall time consumption (including computation and communication) in the $l$-th training round. $B_{max}^{(i)}$ is the largest local batch size permitted to the $i$-th client with the consideration of the client's memory limit. Notably, the local batch size of each client is adjusted in every training round due to the dynamic network condition. Last but not least, the adjustment of the local batch size for each client cannot change the final model weights of each layer on each client, thanks to the weight-invariant property. Therefore, the introduction of the adaptive batch size algorithm does not influence the final model performance on each client. The pseudo-code of FedACnnL with the adaptive batch size algorithm is shown in the algorithm \ref{al1}.
\begin{algorithm2e}[!tb]
\caption{FedACnnL with the adaptive batch size algorithm}\label{al1}
\LinesNumbered
\SetKwData{Left}{left}\SetKwData{This}{this}\SetKwData{Up}{up}
\SetKwFunction{Union}{Union}\SetKwFunction{FindCompress}{FindCompress}
\SetKwInOut{Input}{Input}\SetKwInOut{Output}{Output}
\Input{The participating clients, the regularization coefficient $\gamma$, the initial local batch size $B_1^{(i)}$, the random seed $s$;}
\Output{A global trained model $\{\hat{\textbf{W}}_l\}_{l=1}^L$;}
\BlankLine
Client initialization: initialize a local DNN model $\{\textbf{W}_l\}_{l=1}^L$ and generate a same set of label encoding matrices $\mathcal{Q}=\{\textbf{Q}_l\}_{l=1}^{L-1}$ by $s$;\\
\For{$l=1~to~L$}{
\For{Clients $i=1~to~C$ in parallel}{Initialize $\textbf{X}_l^{(i)T}\textbf{X}_l^{(i)}$ and $\textbf{X}_l^{(i)T}\bar{\textbf{Z}}_l^{(i)}$ respectively with zero matrices;\\
\For{batch of samples \{$\textbf{X}^{\{b\}},\textbf{Y}^{\{b\}}$\} with the local batch size $B_{l}^{(i)}$ sampled from $\mathcal{D}_i$ without replacement}{
$\textbf{X}_1^{\{b\}}=\textbf{X}^{\{b\}}$; $\bar{\textbf{Z}}_l^{\{b\}}=\textbf{Y}^{\{b\}}$;\\
\If{$l>1$}{$\textbf{X}_l^{\{b\}}={f_{l-1}(\underbrace{\cdots{f_1(\textbf{X}_1^{\{b\}}\hat{\textbf{W}}_1)\cdots}}_{\textbf{X}_{l-1}^{\{b\}}}\hat{\textbf{W}}_{l-1})}$;\\
\If{$l<L$}{
$\bar{\textbf{Z}}_l^{\{b\}}=\bar{\textbf{Z}}_l^{\{b\}}\textbf{Q}_l$;}}
$\textbf{X}_l^{(i)T}\textbf{X}_l^{(i)}+=\textbf{X}_l^{\{b\}T}\textbf{X}_l^{\{b\}}$;\\ $\textbf{X}_l^{(i)T}\bar{\textbf{Z}}_l^{(i)}+=\textbf{X}_l^{\{b\}T}\bar{\textbf{Z}}_l^{\{b\}}$\;
}
Upload $\textbf{X}_l^{(i)T}\textbf{X}_l^{(i)}$ and $\textbf{X}_l^{(i)T}\bar{\textbf{Z}}_l^{(i)}$ to the PS\;
}
The PS \textbf{does}:\\
Record the item arrival time $T_l^{(i)}$ of each client;\\
Generate $B_{l+1}^{(i)}$ using the equation \ref{eq7} for each client;\\
$\textbf{A}_0=\sum_{i=1}^C{\textbf{X}_l^{(i)T}\textbf{X}_l^{(i)}}$; $\textbf{A}_1=\sum_{i=1}^{C}\textbf{X}_l^{(i)T}\bar{\textbf{Z}}_l^{(i)}$;\\
Generate $\hat{\textbf{W}}_l$ using the equation \ref{eq5};\\
Send $\hat{\textbf{W}}_l$ and $B_{l+1}^{(i)}$ back to each client; \\
Each client \textbf{does}:\\
Replace $\textbf{W}_l$ with $\hat{\textbf{W}}_l$;\\
Set the local batch size to $B_{l+1}^{(i)}$;
}
\end{algorithm2e}
\subsection{Theoretical Analysis on Computation and Communication}\label{p3p3}
To prove the superiority of our proposed gradient-free framework (i.e. FedACnnL) on the training time, we compare the computational and communication complexity of it with other gradient-free frameworks (i.e. the conventional zeroth-order frameworks: FedZO \cite{fang2022communication} and BAFFLE \cite{feng2025baffle}) theoretically. For a fair comparison, we assume that the local number of samples and the local batch size on each client are equal and fixed (i.e. $|\mathcal{D}_i|=|\mathcal{D}|$, $B_l^{(i)}=B$) during the whole FL process of each framework.

\textbf{Computational complexity.} In the conventional paradigm of FL, the whole training process contains a number of training rounds and each training round requires each client to conduct several repeated visits (i.e. local epochs) on its whole local dataset. Meanwhile, each epoch includes multiple iterations, in which each client samples a batch of data and labels from $\mathcal{D}_i$ without replacement and conducts certain computations on the batch of data and labels. Since each DNN layer in FedACnnL converges in one training round which contains one local epoch, FedACnnL only requires $L$ training rounds in total and each training round contains one local epoch. That is greatly smaller than both FedZO and BAFFLE, in which 
more than 20/40 training rounds (each contains more than one local epoch) are required to achieve a certain model performance empirically on the MNIST \cite{li_deng_mnist_2012}/CIFAR-10 \cite{krizhevsky2009learning} dataset using LeNet-5 (5 layers) \cite{lecun1998gradient}. Therefore, our analysis focuses on the computations conducted on the batch of data and labels to prove a lower batch-level computational complexity for FedACnnL. In FedACnnL, the batch-level computations contain one forward pass and three matrix multiplications (line 8 and line 10 $\sim$ 14 in the algorithm \ref{al1}). However, in FedZO and BAFFLE, the batch-level computations incorporate performing a large number of finite difference processes to accurately approximate the gradient. In each finite difference process, two forward passes should be conducted. After removing the computational complexity of one forward pass for each framework, the complexity of the remaining three matrix multiplications in FedACnnL is $\mathcal{O}(B(d_yd_l+d_{l-1}^2+d_{l-1}d_l))$ in the arbitrary $l$-th training round. Meanwhile, more than one forward passes exists in FedZO or BAFFLE and each of them has a complexity of $\mathcal{O}(\sum_{l=1}^LBd_{l-1}d_l)$. Thus, even if FedZO or BAFFLE merely conducts the finite difference process once (which is impractical) in the batch-level computations, FedACnnL can still save a considerable amount of training time. This theoretical conclusion is experimentally verified in the section \ref{p5p2}.

\textbf{Communication complexity}. In each training round, each client in FedACnnL uploads the resultant two items (i.e. $\textbf{X}_l^{(i)T}\textbf{X}_l^{(i)}$ and $\textbf{X}_l^{(i)T}\bar{\textbf{Z}}_l^{(i)}$) to the PS and receive the trained global model weights of the current training layer $\hat{\textbf{W}}_l$ from the PS. Notably, $B_l^{(i)}$ is too small and we ignore it. Thus, the total communication complexity for FedACnnL in the arbitrary $l$-th training round is $\mathcal{O}(d_{l-1}^2+d_{l-1}d_l+\theta_l)$, where $\theta_l$ denotes the parameter number of the $l$-th DNN layer. Meanwhile, each client in FedZO or BAFFLE uploads and receives the entire DNN model in each training round, resulting in a communication complexity of $\mathcal{O}(2\sum_{l=1}^{L}\theta_l)$ per round. Whether in the convolutional layer or the fully-connected layer, $\theta_l$ is equal to $d_{l-1}d_l$. Therefore, FedACnnL can achieve a lower communication complexity than FedZO and BAFFLE unless the output dimensionality of a certain layer exceeds the parameter number of the entire DNN model. That is a rare case. Furthermore, since FedACnnL requires only $L$ training rounds while FedZO and BAFFLE need a great number one experimentally (see section \ref{p5p2}), the total communication complexity of FedACnnL in the whole FL process is much lower than FedZO and BAFFLE.

\section{Design of pFedACnnL}
\subsection{Overview of pFedACnnL}
\begin{figure*}[tbp]
\centering
\includegraphics[width=16cm]{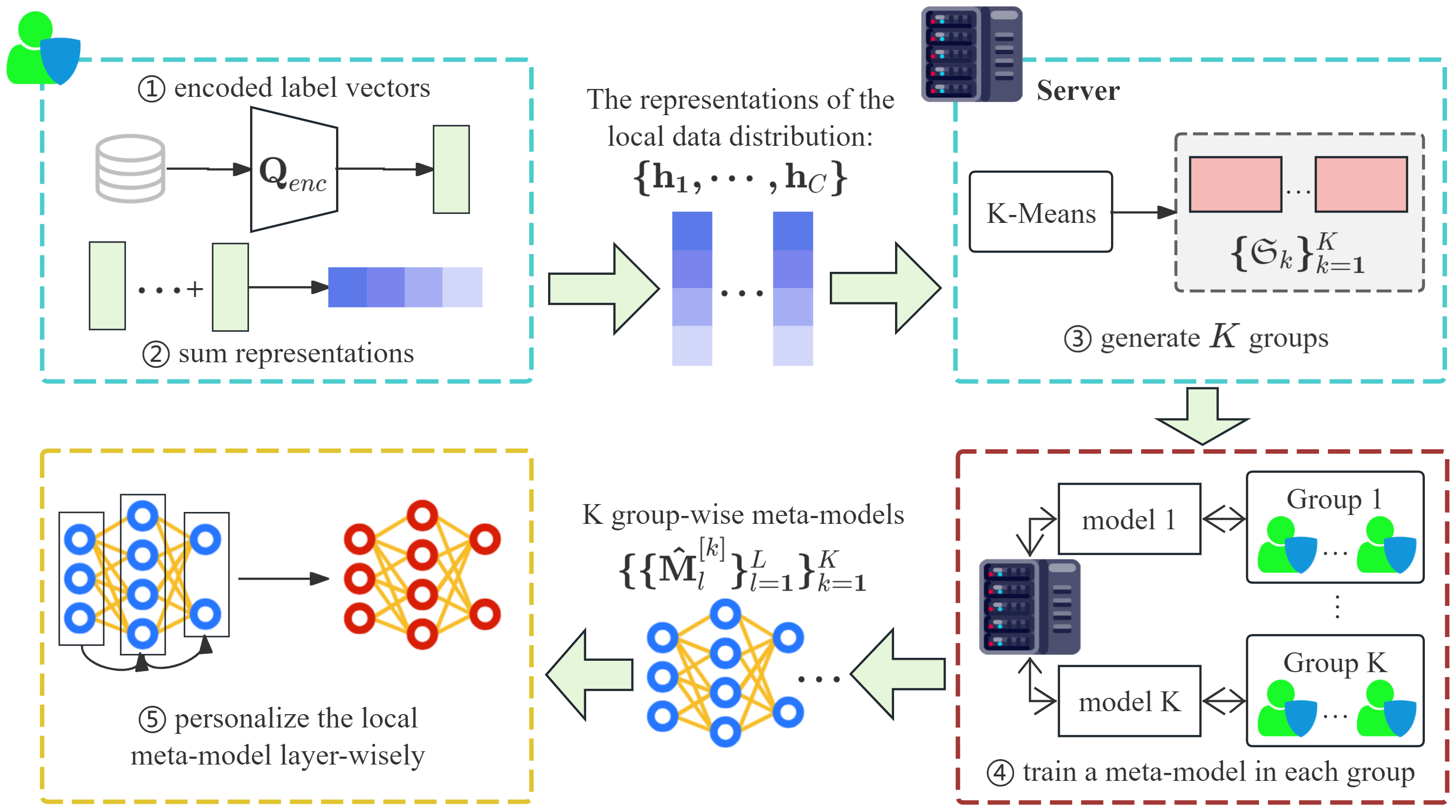}%
\caption{The overview of pFedACnnL. The cyan dashed box represents the initialization stage while the brown dashed box and the orange dashed box represent the federated optimization stage and the local personalization stage respectively.}
\label{fig3}
\end{figure*}

To tackle the heterogeneous data distribution problem that restricts the single global model from performing well on each client's task, we propose pFedACnnL, a personalized version of FedACnnL. pFedACnnL encompasses three stages: the initialization stage, the federated optimization stage and the local personalization stage. An overview of pFedACnnL is illustrated in Fig. \ref{fig3}. 

\textbf{The initialization stage.} In the same way as FedACnnL, each client utilizes the random seed $s$ received from the PS to generate an identical set of label encoding matrices $\mathcal{Q}$. Then, a local model is initialized with the same structure on each client. Next, different from FedACnnL, pFedACnnL requires each client to upload an encoded label vector, which is the representation of the local data distribution on each client, to the PS. Finally, the PS uses these encoded label vectors to conduct the $K$-means clustering algorithm and classifies the clients into $K$ groups. In this way, pFedACnnL enables the clients that have similar local data distribution to train a joint meta-model, namely the group-wise meta-model, in the next stage. By doing so, each group-wise meta-model can contain a specialized unique common knowledge within its group and make it easier for its clients to conduct local personalization than a single global model (see section \ref{p5p3}).

\textbf{The federated optimization stage.} In each group, the clients follow the standard process of FedACnnL. Specifically, in the arbitrary $l$-th federated training round, each client conducts the local computation and uploads the resultant two items (i.e. $\textbf{X}_l^{(i)T}\textbf{X}_l^{(i)}$ and $\textbf{X}_l^{(i)T}\bar{\textbf{Z}}_l^{(i)}$) to the PS as in FedACnnL. Next, the PS aggregates the items respectively by a sum operation within each group and then generates the model weights of the $l$-th layer for the $K$ group-wise meta-models respectively. At the end of the federated training round, each client receives the model weights belonging to its group to update the $l$-th layer of the local model. After pFedACnnL completes its all federated training rounds, $K$ group-wise meta-models are born across the clients.

\textbf{The local personalization Stage.} Each client uses the local meta-model to achieve a personalized model catering for the local data distribution by using a gradient-free meta-learning method. This method follows a similar optimization process to ACnnL but updates the model weights of each layer using closed-from LS solutions of a local PFL objective. Specifically, in the arbitrary $l$-th local personalization iteration, each client generates its pseudo label matrix and real input matrix of the $l$-th hidden layer as in ACnnL. Then, a local PFL objective is designed to achieve supervised learning of the $l$-th hidden layer and meta-learning from the group-wise meta-model simultaneously. The personalized model weights of the $l$-th hidden layer are obtained by analytically solving the objective on each client. After that, each client computes the real input matrix of the next layer and steps into personalizing the next layer. The local personalization stage finishes when all clients have personalized each layer of their local meta-models.

\subsection{The Initialization Stage}
After each client initializes the local model and the set of label encoding matrices $\mathcal{Q}$ in the same way as FedACnnL, it uploads an encoded label vector to the PS for conducting the $K$-means clustering algorithm. Specifically, the encoded label vector indicates the information of the local data distribution. This information can be directly represented by the sum of all local label vectors, i.e. $\sum_{\textbf{y}_n\in{\mathcal{D}_i}}\textbf{y}_n$ for the arbitrary $i$-th client. However, the sum of the label vectors can reveal the client's preference or privacy. Therefore, to safeguard the privacy to some extent and reduce the communication burden, the sum of the label vectors should be normalized and then encoded by $\textbf{Q}_{enc}$ which is selected from $\mathcal{Q}$ and satisfies the following equation:
\begin{equation}
\textbf{Q}_{enc}=\underset{\textbf{Q}_l\in{\mathcal{Q}}}{\arg\min}~d_l
\label{eq8}
\end{equation}
Then, the uploaded encoded vector for the arbitrary $i$-th client (i.e. $\textbf{h}_i$) can be mathematically represented as:
\begin{equation}
\textbf{h}_i=\textbf{Q}_{enc}\frac{1}{|\mathcal{D}_i|}\sum_{\textbf{y}_n\in{\mathcal{D}_i}}\textbf{y}_n=\frac{1}{|\mathcal{D}_i|}\sum_{\textbf{y}_n\in{\mathcal{D}_i}}\textbf{Q}_{enc}\textbf{y}_n
\label{eq9}
\end{equation}
After the PS receives the encoded vectors $\{\textbf{h}_i\}_{i=1}^C$ from all clients, it utilizes these vectors to conduct the $K$-means clustering algorithm to classify the clients into $K$ groups, i.e. $\{\mathfrak{S}_k\}_{k=1}^{K}$, where $k$ is the index of a group and $\mathfrak{S}_k$ denotes a sequence of clients' indices. As a result, the clients belonging to the same group have similar local data distribution, resulting in a potential common knowledge within each group to be tapped in the following federated optimization stage.

\subsection{The Federated Optimization Stage}
In this stage, the clients belonging to the same group collaboratively train a joint meta-model by following the standard process of FedACnnL. Specifically,
in the arbitrary $l$-th federated training round of the $k$-th group, pFedACnnL reformulates the federated optimization problem in the $l$-th layer of the group-wise meta-model by reconstructing the distributed LS problem in the equation \ref{eq3} as follows.
\begin{equation}
    \begin{split}
    &\underset{\textbf{M}_l^{[k]}}{\arg\min}~\sum_{i\in{\mathfrak{S}_k}}F(\textbf{M}_l^{[k]};\mathcal{D}_i) \\
    where~F(\textbf{M}_l^{[k]}&;\mathcal{D}_i)=\Vert{\bar{\textbf{Z}}_l^{(i)}-\textbf{X}_l^{(i)}\textbf{M}_l^{[k]}}\Vert_2^2+\gamma\Vert{\textbf{M}_l^{[k]}}\Vert_2^2  \\
    \end{split}
\label{eq10}
\end{equation}
where $\textbf{M}_l^{[k]}$ denotes the meta-model weights of the $l$-th layer in the $k$-th group. Then, we can present the closed-form LS solutions of the equation \ref{eq10} by putting its derivatives with respect to $\textbf{M}_l^{[k]}$ to 0:
\begin{equation}
\hat{\textbf{M}}_l^{[k]} = (\underbrace{\sum_{i\in{\mathfrak{S}_k}}{\textbf{X}_l^{(i)T}\textbf{X}_l^{(i)}}}_{\textbf{A}_0^{[k]}}+\gamma{\textbf{I}})^{-1}(\underbrace{\sum_{i\in{\mathfrak{S}_k}}\textbf{X}_l^{(i)T}\bar{\textbf{Z}}_l^{(i)}}_{\textbf{A}_1^{[k]}})
\label{eq11}
\end{equation}
where $\hat{\textbf{M}}_l^{[k]}$ denotes the trained meta-model weights of the $l$-th layer in the $k$-th group. $\textbf{A}_0^{[k]}$ and $\textbf{A}_1^{[k]}$ are the aggregated results of the resultant two items (i.e. $\textbf{X}_l^{(i)T}\textbf{X}_l^{(i)}$ and $\textbf{X}_l^{(i)T}\bar{\textbf{Z}}_l^{(i)}$) uploaded by the clients in the $k$-th group. Based on the reformulation, we re-describe the federated optimization stage elaborately. In the arbitrary $l$-th federated training round, each client conducts its local computation as in FedACnnL and uploads the resultant two items to the PS. Then, the PS generates the aggregated results for each group respectively, i.e. $\{\textbf{A}_0^{[k]},\textbf{A}_1^{[k]}\}_{k=1}^K$ and utilizes them to obtain $\{\hat{\textbf{M}}_l^{[k]}\}_{k=1}^K$ for each group, by using the equation \ref{eq11}. At the end of the federated training round, each client receives the trained meta-model weights of the $l$-th layer belonging to its group and uses them to replace the local model weights of the $l$-th layer. The federated optimization stage completes when all layers of the local model on each client have been replaced once. By the way, the adaptive batch size algorithm is also applied in pFedACnnL. Specifically, before the PS aggregates the resultant two items in each federated training round, it records the item arrival times for all clients and calculates the local batch size of the next round for each client by using the equation \ref{eq7}. At the end of each federated training round, the PS sends the calculated result accompanied with the trained meta-model weights back to each client.

\subsection{The Local Personalization Stage}
After the federated optimization stage, each client obtains a meta-model that incorporates the common knowledge applicable to all clients' tasks within its group. The aim of the local personalization stage on each client is to generate a personalized model from its meta-model to promote the model performance on each client's task. For a fast generation in our gradient-free setting, a gradient-free meta-learning method is proposed and incorporated in this stage. Specifically, each client personalizes and updates the local model weights in a layer-wise way like ACnnL. In the arbitrary $l$-th local personalization iteration of the $i$-th client, the client gets its pseudo label matrix $\bar{\textbf{Z}}_l^{(i)}$ and real input matrix $\textbf{X}_l^{(i)}$ as in ACnnL. Then, pFedACnnL formulates the personalization of the $l$-th layer on the $i$-th client as an LS problem with the following PFL objective:
\begin{equation}
\underset{\textbf{W}_l^{(i)}}{\arg\min}~\underbrace{\Vert{\bar{\textbf{Z}}_l^{(i)}-\textbf{X}_l^{(i)}\textbf{W}_l^{(i)}}\Vert_2^2}_{g(\cdot)}+\epsilon\underbrace{\Vert{\textbf{W}_l^{(i)}-\hat{\textbf{M}}_l^{[k]}}\Vert_2^2}_{h(\cdot)}
\label{eq12}
\end{equation}
where $\textbf{W}_l^{(i)}$ denotes the model weights of the $l$-th layer on the $i$-th client. $g(\cdot)$ denotes the supervised learning of the $l$-th hidden layer while $h(\cdot)$ denotes the meta-learning from the client's meta-model. $\epsilon$ is the hyperparameter that controls the tradeoff between the personalization performance and the generalization performance in FL. By putting the derivatives of the equation \ref{eq12} with respect to $\textbf{W}_l^{(i)}$ to 0, we can get the personalized model weights of the $l$-th layer on the $i$-th client, i.e. $\hat{\textbf{W}}_l^{(i)}$, analytically as:
\begin{equation}
\hat{\textbf{W}}_l^{(i)} = ({\textbf{X}_l^{(i)T}\textbf{X}_l^{(i)}}+\epsilon{\textbf{I}})^{-1}(\textbf{X}_l^{(i)T}\bar{\textbf{Z}}_l^{(i)}+\epsilon{\hat{\textbf{M}}_l^{[k]}})
\label{eq13}
\end{equation}
After replacing the local model weights with the personalized ones, each client computes the real input matrix of the next layer, i.e. $\textbf{X}_{l+1}^{(i)}=f_l(\textbf{X}_l^{(i)}\hat{\textbf{W}}_l^{(i)})~or~f_l(\textbf{X}_l^{(i)}{\star}\hat{\mathcal{W}}_l^{(i)})$, and steps into the next local personalization iteration. The local personalization stage finishes when all layers of the local meta-model have been personalized.

\section{Performance Evaluation}\label{p5}

\subsection{Experimental Settings}\label{p5p1}
\textbf{Testbed.} We implement a prototype of our proposed frameworks on a CPU-cluster-based testbed with 101 nodes. The first node is treated as the PS while the other nodes is regarded as the clients. Each node is equipped with four Intel(R) Xeon(R) Platinum 9242 CPUs. We implement one process on each node with a restriction on the available number of CPU cores. The available number of CPU cores on each node is set randomly ranging from 1 to 4 to emulate the edge devices that have different computing power and network conditions. The nodes are connected by InfiniBand network with a bandwidth of 100Gbps.

\textbf{Datasets.} We consider classification tasks using three datasets: the MNIST dataset \cite{li_deng_mnist_2012}, the synthetic dataset \cite{li2020federated} and the CIFAR-10 dataset \cite{krizhevsky2009learning} to simulate three distinct levels of task difficulty in FL (from easy to hard). The MNIST dataset is a handwritten digit dataset with 70,000 instances, in which each sample contains a binary image with a resolution of $28\times{28}\times1$ and a label indicator ranging from 0 to 9. The synthetic dataset is produced artificially following the methodology outlined
in \cite{li2020federated}, integrating two parameters, $\bar{\alpha}$ and $\bar{\beta}$, to govern the degree of
distinction between the local model and dataset for every individual
client. The CIFAR-10 dataset contains 60,000 tiny RGB images, which have a resolution of $32\times{32}\times3$ and can be classified into 10 categories. For the MNIST and CIFAR-10 datasets, we employ the Dirichlet sampling $p_{u,i}{\sim}Dir_N(\beta)$ to generate the heterogeneous sample partitions of the datasets over 100 clients, where $p_{u,i}$ denotes the probability of allocating a sample of label $u$ to the $i$-th client. A smaller $\beta$  means more diversity in the data distribution between local clients. We set $\beta=0.1$ to emulate the highly heterogeneous data environment \cite{zhang2025lcfed}. For the synthetic dataset, we adhere the power law distribution to distribute the samples
among 100 clients. Meanwhile, we set $\bar{\alpha}$ and $\bar{\beta}$ as 0.5 as in \cite{chen2024gradient}.

\textbf{Baselines and Metircs.} To prove the superiority of our proposed AFL 
 (i.e. gradient-free FL) framework, FedACnnL, on the efficiency of training DNN, we compare it with the following four frameworks (including GD-based and gradient-free frameworks) and choose the averaged total training time of each client as the main metric. The total training time of each client consists of the computation time, the idle time and the computation time. The computation time is the time cost for conducting local computation. The idle time for a client is the time span during which it awaits the completion of the local computation by the client with the poorest computing power. The communication time is the time expense of transmitting data between the client and the PS.

 \begin{itemize}
    \item FedACnnL-WA: This is the version of FedACnnL without the adaptive batch size algorithm. The existence of it is to evaluate the effectiveness of the introduced adaptive batch size algorithm to our AFL framework.
    \item FedAvg \cite{mcmahan_communication-efcient_nodate}: It is the classical first-order FL framework that enables multiple clients collaboratively to train a high-quality global model.
    \item FedZO \cite{fang2022communication}: It is the standard zeroth-order FL framework that enables each client to approximate the local gradients with only DNN inference processes by using the finite difference technique.
    \item BAFFLE \cite{feng2025baffle}: A novel baseline for zeroth-order FL frameworks, which resorts to the finite difference technique to approximate the gradients locally on each client.
\end{itemize}
Furthermore, to evaluate the effectiveness of our proposed APFML framework, pFedACnnL (the personalized version of FedACnnL), we compare it with the vanilla FedACnnL and another four PFL frameworks. We choose the overall test accuracy of all clients as the metric of the model performance.
\begin{itemize}
    \item FedACnnL: The vanilla version of our proposed AFL framework, which is directly used to prove the benefits of its personalized version on the model performance.
    \item pFedACnnL-WC: This is the version of pFedACnnL without the $K$-means clustering algorithm. In other words, the federated optimization stage completely follows the standard process of FedACnnL without any in-group aggregation. The existence of it is to prove the necessity of the component (i.e. the $K$-means clustering algorithm) in pFedACnnL.
    \item pFedACnnL-WP: This is the version of pFedACnnL without the local personalization stage. After the federated optimization stage, the framework kills the process. The existence of it is to prove the necessity of the component (i.e. the local personalization stage) in pFedACnnL.
    \item pFedMe \cite{t2020personalized}: A first-order PFL framework that utilizes Moreau envelops, which is a meta-learning technique to find a balance between generalization and personalization performance in FL.
    \item pFedZO \cite{chen2024gradient}: A zeroth-order PFL framework that extends pFedMe to FedZO by replacing the backpropagation for gradient computation with the finite difference technique for gradient approximation.
\end{itemize}

\textbf{Models.} As in \cite{chen2024gradient}, we implement our frameworks under both convex and non-convex settings. For the convex setting, we consider a multinomial logistic regression (LR) model with softmax activation. For the non-convex setting, a DNN with a structure of two fully-connected hidden layers is implemented with hidden layer sizes of [128,64] and ReLU activation functions. Since the designs of FedACnnL and pFedACnnL involve no utilization of any loss function, it is unnecessary to specify a loss function for FedACnnL, pFedACnnL and their variants while all other frameworks use cross-entropy as the loss function. Furthermore, to evaluate the capability of training a deeper and wider DNN model in the complex machine learning task (i.e. the CIFAR-10 dataset), a 5-layer deep convolutional neural network (DCNN) with a structure of $\{[\textbf{Conv}(5\times{5})\times512-\textbf{LeakyReLU}]-[\textbf{AvgPool}(2\times2)]-[\textbf{Conv}(3\times{3})\times1024-\textbf{LeakyReLU}]-[\textbf{AvgPool}(2\times2)]-[\textbf{Conv}(3\times{3})\times2048-\textbf{LeakyReLU}]-[\textbf{Conv}(3\times{3})\times2048-\textbf{LeakyReLU}]-[\textbf{MLP}(10)]\}$ \cite{zhuang2025analytic} is implemented to FedACnnL, pFedACnnL and their variants. 

\textbf{Hyperparameter setting.} We choose $\gamma=100$, $\epsilon=2500$ and $K=10$ for FedACnnL and pFedACnnL since this combination achieves the best model performance empirically in our experiments. In our first experiment, in which we compare FedACnnL with other frameworks in the averaged total training time, the number of training rounds is set to 20 on the MNIST dataset according to \cite{feng2025baffle}, while it is set to 600 on the synthetic dataset according to \cite{t2020personalized}. In the second experiment, in which we evaluate our APFML framework, pFedACnnL and its variants choose the test accuracy after they finish their all stages as the final model performance. Meanwhile, all other frameworks choose the best test accuracy within 1,000 training rounds as the final model performance, just as in \cite{chen2024gradient}.

\begin{figure*}[!tbp]
\centering
\includegraphics[width=16cm]{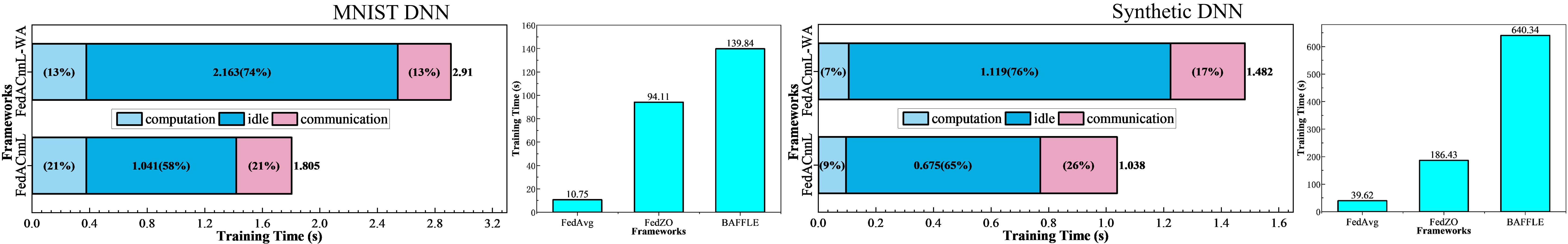}%
\caption{The averaged total training time of each client in each framework on the MNIST and synthetic datasets.}
\label{fig5}
\end{figure*}

\begin{figure}[!tbp]
\centering
\includegraphics[width=8cm]{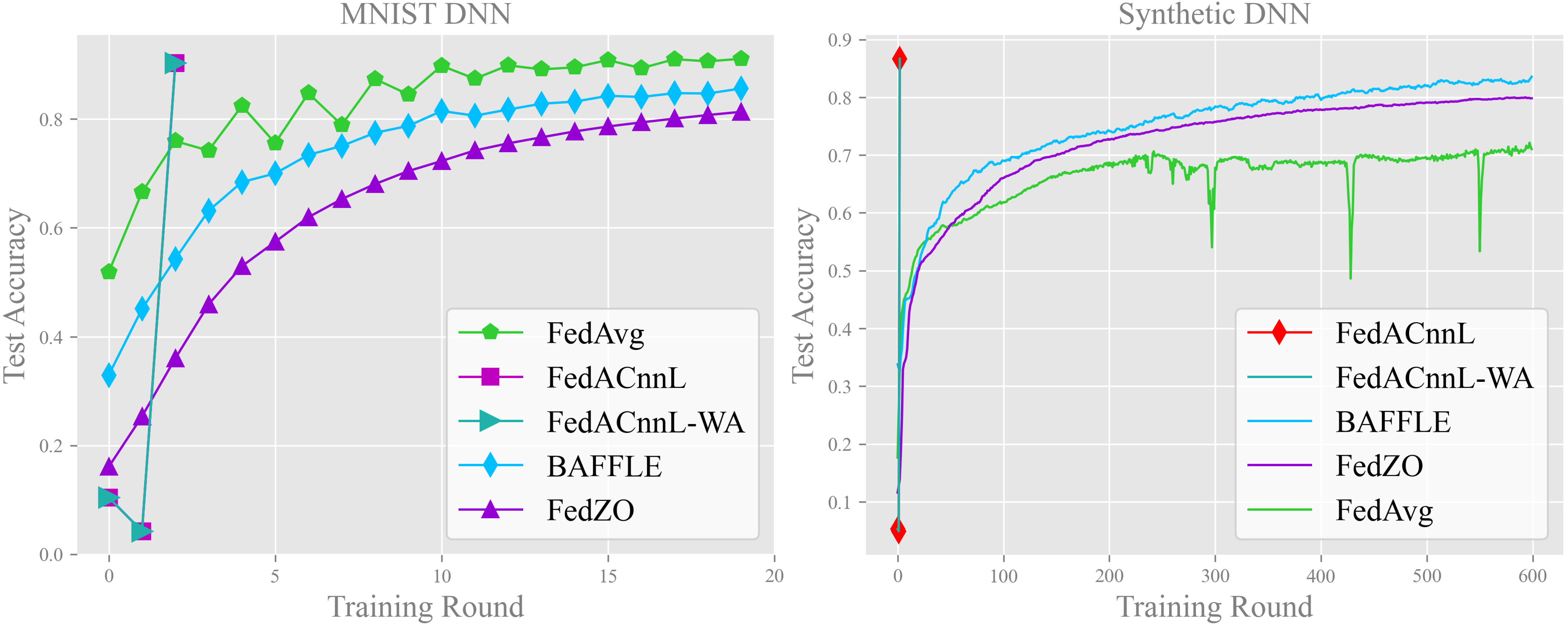}%
\caption{The overall test accuracy of all clients versus training round curve during the training procedure on the MNIST and synthetic datasets.}
\label{fig4}
\end{figure}

\subsection{Training Efficiency}\label{p5p2}
To highlight the efficiency of FedACnnL on training DNN, we carry out the experiment in the non-convex setting to validate the training time reduction of FedACnnL over the aforementioned first- and zeroth-order FL frameworks (i.e. FedAvg, FedACnnL-WA, FedZO and BAFFLE). Before we discuss the training time, we should establish the model performance guarantee for FedACnnL. Fig. \ref{fig4} demonstrates the overall test accuracy of all clients versus training round curve during the training procedure. It shows that FedACnnL can approach a competitive model performance with the first-order FL framework FedAvg on the MNIST dataset while it achieves the state-of-the-art (SOTA) model performance on the synthetic dataset. Meanwhile, FedACnnL-WA achieves the same model performance as FedACnnL due to the weight-invariant property which has been discussed in the section \ref{p3p1}. On the basis of the performance superiority, we present the averaged total training time for each framework in Fig. \ref{fig5}. As illustrated in Fig. \ref{fig5}, FedACnnL can reduce 83\%, 98\% and 99\% training time of FedAvg, FedZO and BAFFLE respectively while the reduction ratio is more than 97\% on the synthetic dataset. That is because FedACnnL updates its model weights only once in a layer-wise manner while the other frameworks require clients to iteratively update their models locally and need substantial training rounds. Furthermore, the total time per training round in FedZO and BAFFLE is $4.7s, 7.0s$ on the MNIST dataset and $0.31s,1.07s$ on the synthetic dataset while that in FedACnnL is $0.60s$ and $0.35s$ on the MNIST and synthetic datasets. This fact validates our theoretical analysis in the section \ref{p3p3} which indicates that FedACnnL has a lower complexity of computation and communication on each client in each training round than FedZO and BAFFLE. Finally, FedACnnL reduces $52\%$ and $40\%$ idle time of FedACnnL-WA on the MNIST and synthetic dataset, proving the effectiveness of the adaptive batch size algorithm in which the time expenses of the local computation for the lagging clients are reduced.

\begin{figure}[!tbp]
\centering
\includegraphics[width=8cm]{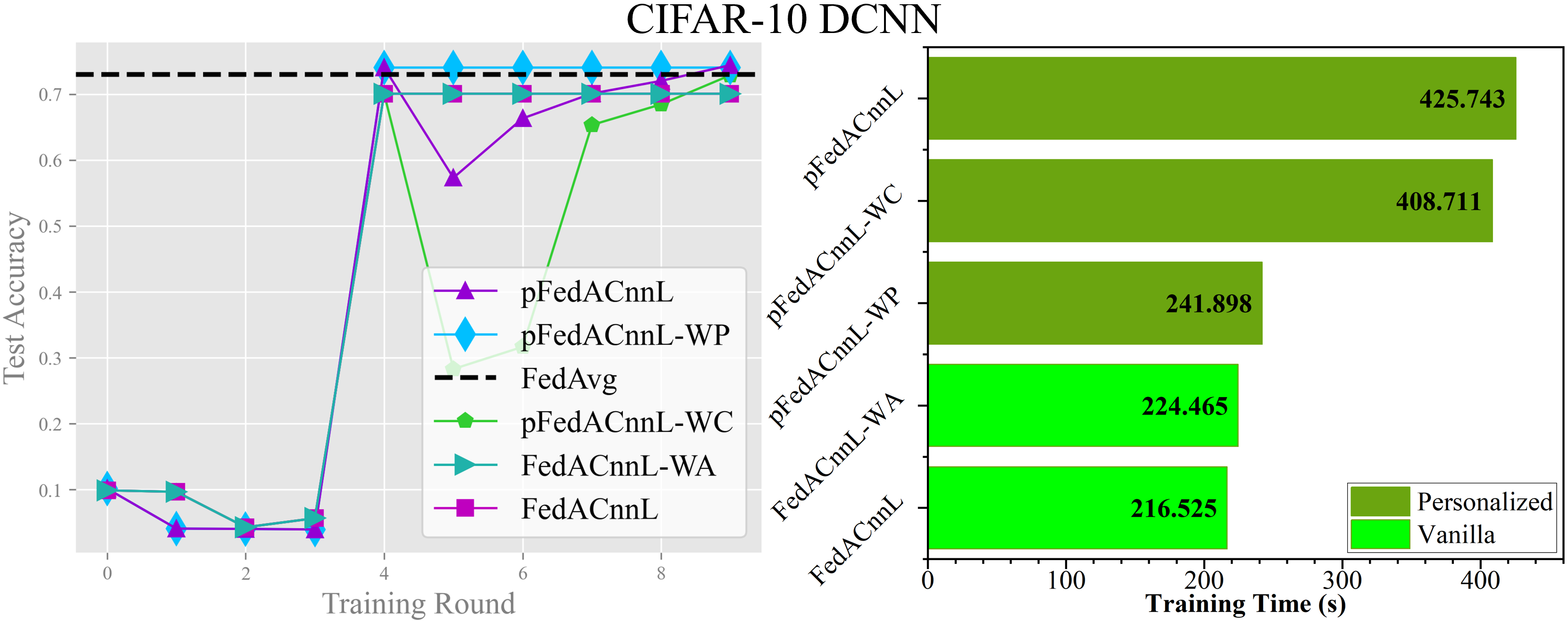}%
\caption{The model performance and training time of our proposed frameworks on the CIFAR-10 dataset using the DCNN model.}
\label{fig6}
\end{figure}

\textbf{Performance on a deeper and wider DNN model}. To further prove the effectiveness of our proposed frameworks on the DNN training, we implement FedACnnL, pFedACnnL and their variants on training the aforementioned DCNN model. Fig \ref{fig6} shows the model performance and training time of our proposed frameworks on the CIFAR-10 dataset using the DCNN model. Since pFedACnnL-WP, FedACnnL-WA and FedACnnL do not have the local personalization stage, we fill in the test accuracy of the blank training rounds (i.e. rounds from 5 to 9), which are used for local personalization in pFedACnnL and pFedACnnL-WC, for them with their final test accuracy. FedACnnL can approach a competitive model performance with FedAvg (70.06\% versus 73\%) while its personalized version i.e. pFedACnnL-WP and pFedACnnL can achieve a test accuracy of 74\% and 74.40\%, which is 1\% better than FedAvg. Furthermore, all of our proposed frameworks can finish their training in a minute-level total training time.

\begin{table*}[!tbp]
\renewcommand{\arraystretch}{1.3}
\setlength\tabcolsep{4.0pt}
\caption{Table of model performance.}
\begin{center}
\centering
\begin{tabular}{c|c|cccccc}
%添加顶部横线 
\Xhline{1.5 pt}
%输入标题
\multirow{2}{*}{\textbf{Settings}}&\multirow{2}{*}{\textbf{Dataset-Model}} & \multicolumn{6}{c}
{\textbf{Frameworks}}\\
\cline{3-8}
&&FedACnnL&pFedACnnL-WC&pFedACnnL-WP&pFedMe&pFedZO&pFedACnnL\\
\cline{1-8}
\multirow{2}{*}{\textbf{non-convex}}&MNIST-DNN&$90.76(\pm0.21)$&$96.85(\pm0.04)$&$93.66(\pm0.07)$&$\textbf{96.92}(\pm0.03$)&$96.17(\pm0.08)$&$96.23(\pm0.01)$\\
&Synthetic-DNN&$86.09(\pm0.06)$&$88.63(\pm0.01)$&$89.22(\pm0.03)$&$88.58(\pm0.17)$&$89.05(\pm0.04)$&$\textbf{90.00}(\pm0.03)$\\
\cline{1-8}
\multirow{2}{*}{\textbf{convex}}&MNIST-LR&$87.11(\pm0.00)$&$\textbf{96.24}(\pm0.00)$&$91.79(\pm0.01)$&$93.22(\pm0.03)$&$93.09(\pm0.05)$&$95.81(\pm0.01)$\\
&Synthetic-LR&$81.80(\pm0.00)$&$87.81(\pm0.00)$&$88.33(\pm0.01)$&$83.13(\pm0.01)$&$72.14(\pm0.08)$&$\textbf{88.96}(\pm0.01)$\\

% % \cite{2-3}
% % \cline{2-7}
% \Xhline{1.5 pt}
% %添加标题和内容之间的横线
% % \Xhline{0.5 pt}
% % \centering
% \cline{2-5}
% &Art&Clipart&Product&Real World&\\
% % \cline{2-4}
% \Xhline{0.5 pt}
% \textbf{FedMPI}&\textbf{61.65($\uparrow$6.09)}&70.86&82.11&76.69&\textbf{74.43 ($\uparrow$4.75})\\
% FedZO&54.12&59.62&75.52&72.64&66.99\\
% $K$ 100&53.05&56.95&74.01&68.79&64.56\\
% $K$ 200&53.41&58.67&75.89&71.68&66.45\\
% $K$ 500&55.56&62.29&79.10&75.14&69.68\\
% \Xhline{0.5 pt}
% FedAvg&59.14&75.05&85.12&80.15&76.97\\
% \textbf{pFedMPI}&57.71&74.86&\textbf{87.95($\uparrow$2.83)}&79.58&\textbf{77.35($\uparrow$0.42)}\\

\Xhline{1.5 pt}
\end{tabular}
\label{tab2}
\end{center}
\end{table*}

\subsection{Personalized Model Performance}\label{p5p3}
We evaluate our proposed APFML framework in both the convex and non-convex settings. The evaluation is done by taking the mean and standard deviation of the test accuracy from running the experiment for 5 times. Table \ref{tab2} lists the model performance of pFedACnnL, the vanilla FedACnnL and another four PFL frameworks. In the table, we can find that pFedACnnL experiences a performance improvement of $4\%\sim6\%$ in the test accuracy under the non-convex setting while it has an improvement of $7\%\sim8\%$ under the convex setting, compared to the vanilla FedACnnL. Furthermore, as illustrated in Fig. \ref{fig6} and Table \ref{tab2}, in most cases (except in the MNIST dataset), pFedACnnL has a better model than its versions without a certain component (i.e. pFedACnnL-WC and pFedACnnL-WP). For instance, in Fig. \ref{fig6}, pFedACnnL achieves a test accuracy of 74.40\% on the CIFAR-10 dataset while pFedACnnL-WC and pFedACnnL-WP achieve a test accuracy of 72.88\% and 74.03\% respectively. Notably, on the CIFAR-10 and synthetic datasets, compared with pFedACnnL-WP, pFedACnnL-WC which uses a single global model has an even lower model performance than pFedACnnL, proving the effectiveness of the grouping mechanism in the highly heterogeneous data environment. The special case is that pFedACnnL-WC outperforms pFedACnnL in the MNIST dataset. That is because the task difficulty of the MNIST dataset is comparatively lower than that of the synthetic and CIFAR-10 datasets. Thus, introducing too many complex techniques (including DNN, group-wise meta-learning and etc) to pFedACnnL for easy machine learning tasks can cause an over-fitting problem. Last but not least, pFedACnnL achieves the SOTA performance compared to the first- and zeroth-order PFL frameworks (i.e. pFedMe and pFedZO) in most cases while it achieves a competitive performance with pFedMe (96.23\% versus 96.92\%) in the non-convex setting of the MNIST dataset.

\section{Conclusions}\label{p6}
In this article, we have proposed FedACnnL to enable computation-efficient DNN training in AFL and proposed pFedACnnL based on it to mitigate the heterogeneous data distribution problem in AFL. Specifically, FedACnnL is the federated implementation of ACnnL with an introduction of the adaptive batch size algorithm to boost the training procedure. Meanwhile, pFedACnnL incorporates a gradient-free meta-learning method to generate a tailored model for each client's data distribution. Experimentally, FedACnnL reduces the training time by $83\%\sim99\%$ compared with the conventional first- and zeroth-order FL frameworks while pFedACnnL achieves the SOTA model performance in most cases of convex and non-convex settings.

\bibliographystyle{cas-model2-names}

\bibliography{cas-refs}

\begin{thebibliography}{25}
\expandafter\ifx\csname natexlab\endcsname\relax\def\natexlab#1{#1}\fi
\providecommand{\url}[1]{\texttt{#1}}
\providecommand{\href}[2]{#2}
\providecommand{\path}[1]{#1}
\providecommand{\DOIprefix}{doi:}
\providecommand{\ArXivprefix}{arXiv:}
\providecommand{\URLprefix}{URL: }
\providecommand{\Pubmedprefix}{pmid:}
\providecommand{\doi}[1]{\href{http://dx.doi.org/#1}{\path{#1}}}
\providecommand{\Pubmed}[1]{\href{pmid:#1}{\path{#1}}}
\providecommand{\bibinfo}[2]{#2}
\ifx\xfnm\relax \def\xfnm[#1]{\unskip,\space#1}\fi
%Type = Article
\bibitem[{Ficco et~al.(2024)Ficco, Guerriero, Milite, Palmieri, Pietrantuono and Russo}]{ficco2024federated}
\bibinfo{author}{Ficco, M.}, \bibinfo{author}{Guerriero, A.}, \bibinfo{author}{Milite, E.}, \bibinfo{author}{Palmieri, F.}, \bibinfo{author}{Pietrantuono, R.}, \bibinfo{author}{Russo, S.}, \bibinfo{year}{2024}.
\newblock \bibinfo{title}{Federated learning for iot devices: Enhancing tinyml with on-board training}.
\newblock \bibinfo{journal}{Information Fusion} \bibinfo{volume}{104}, \bibinfo{pages}{102189}.
%Type = Inproceedings
\bibitem[{McMahan et~al.(2017)McMahan, Moore, Ramage, Hampson and y~Arcas}]{mcmahan_communication-efcient_nodate}
\bibinfo{author}{McMahan, B.}, \bibinfo{author}{Moore, E.}, \bibinfo{author}{Ramage, D.}, \bibinfo{author}{Hampson, S.}, \bibinfo{author}{y~Arcas, B.A.}, \bibinfo{year}{2017}.
\newblock \bibinfo{title}{Communication-efficient learning of deep networks from decentralized data}, in: \bibinfo{booktitle}{Artificial intelligence and statistics}, \bibinfo{organization}{PMLR}. pp. \bibinfo{pages}{1273--1282}.
%Type = Article
\bibitem[{Liu et~al.(2024)Liu, Huo, Qu, Xu, Liu, Ma and Huang}]{liu2024fedcd}
\bibinfo{author}{Liu, J.}, \bibinfo{author}{Huo, Y.}, \bibinfo{author}{Qu, P.}, \bibinfo{author}{Xu, S.}, \bibinfo{author}{Liu, Z.}, \bibinfo{author}{Ma, Q.}, \bibinfo{author}{Huang, J.}, \bibinfo{year}{2024}.
\newblock \bibinfo{title}{Fedcd: A hybrid federated learning framework for efficient training with iot devices}.
\newblock \bibinfo{journal}{IEEE Internet of Things Journal} .
%Type = Article
\bibitem[{Sen et~al.(2024)Sen, Qin et~al.}]{sen2024fagh}
\bibinfo{author}{Sen, M.}, \bibinfo{author}{Qin, A.K.}, et~al., \bibinfo{year}{2024}.
\newblock \bibinfo{title}{Fagh: Accelerating federated learning with approximated global hessian}.
\newblock \bibinfo{journal}{arXiv preprint arXiv:2403.11041} .
%Type = Inproceedings
\bibitem[{Feng et~al.(2025)Feng, Pang, Du, Chen, Yan and Lin}]{feng2025baffle}
\bibinfo{author}{Feng, H.}, \bibinfo{author}{Pang, T.}, \bibinfo{author}{Du, C.}, \bibinfo{author}{Chen, W.}, \bibinfo{author}{Yan, S.}, \bibinfo{author}{Lin, M.}, \bibinfo{year}{2025}.
\newblock \bibinfo{title}{Baffle: A baseline of backpropagation-free federated learning}, in: \bibinfo{booktitle}{European Conference on Computer Vision}, \bibinfo{organization}{Springer}. pp. \bibinfo{pages}{89--109}.
%Type = Inproceedings
\bibitem[{Wang et~al.(2019)Wang, Liu, Lin, Lin and Han}]{wang2019haq}
\bibinfo{author}{Wang, K.}, \bibinfo{author}{Liu, Z.}, \bibinfo{author}{Lin, Y.}, \bibinfo{author}{Lin, J.}, \bibinfo{author}{Han, S.}, \bibinfo{year}{2019}.
\newblock \bibinfo{title}{Haq: Hardware-aware automated quantization with mixed precision}, in: \bibinfo{booktitle}{Proceedings of the IEEE/CVF conference on computer vision and pattern recognition}, pp. \bibinfo{pages}{8612--8620}.
%Type = Article
\bibitem[{Dai et~al.(2020)Dai, Low and Jaillet}]{dai2020federated}
\bibinfo{author}{Dai, Z.}, \bibinfo{author}{Low, B.K.H.}, \bibinfo{author}{Jaillet, P.}, \bibinfo{year}{2020}.
\newblock \bibinfo{title}{Federated bayesian optimization via thompson sampling}.
\newblock \bibinfo{journal}{Advances in Neural Information Processing Systems} \bibinfo{volume}{33}, \bibinfo{pages}{9687--9699}.
%Type = Article
\bibitem[{Yi et~al.(2022)Yi, Zhang, Yang and Johansson}]{yi2022zeroth}
\bibinfo{author}{Yi, X.}, \bibinfo{author}{Zhang, S.}, \bibinfo{author}{Yang, T.}, \bibinfo{author}{Johansson, K.H.}, \bibinfo{year}{2022}.
\newblock \bibinfo{title}{Zeroth-order algorithms for stochastic distributed nonconvex optimization}.
\newblock \bibinfo{journal}{Automatica} \bibinfo{volume}{142}, \bibinfo{pages}{110353}.
%Type = Article
\bibitem[{Fang et~al.(2022)Fang, Yu, Jiang, Shi, Jones and Zhou}]{fang2022communication}
\bibinfo{author}{Fang, W.}, \bibinfo{author}{Yu, Z.}, \bibinfo{author}{Jiang, Y.}, \bibinfo{author}{Shi, Y.}, \bibinfo{author}{Jones, C.N.}, \bibinfo{author}{Zhou, Y.}, \bibinfo{year}{2022}.
\newblock \bibinfo{title}{Communication-efficient stochastic zeroth-order optimization for federated learning}.
\newblock \bibinfo{journal}{IEEE Transactions on Signal Processing} \bibinfo{volume}{70}, \bibinfo{pages}{5058--5073}.
%Type = Article
\bibitem[{Chen et~al.(2024)Chen, Chen, Gu and Deng}]{chen2024fine}
\bibinfo{author}{Chen, J.}, \bibinfo{author}{Chen, H.}, \bibinfo{author}{Gu, B.}, \bibinfo{author}{Deng, H.}, \bibinfo{year}{2024}.
\newblock \bibinfo{title}{Fine-grained theoretical analysis of federated zeroth-order optimization}.
\newblock \bibinfo{journal}{Advances in Neural Information Processing Systems} \bibinfo{volume}{36}.
%Type = Article
\bibitem[{Zhuang et~al.(2024)Zhuang, He, Tong, Fang, Sun, Li, Chen and Zeng}]{zhuang2024analytic}
\bibinfo{author}{Zhuang, H.}, \bibinfo{author}{He, R.}, \bibinfo{author}{Tong, K.}, \bibinfo{author}{Fang, D.}, \bibinfo{author}{Sun, H.}, \bibinfo{author}{Li, H.}, \bibinfo{author}{Chen, T.}, \bibinfo{author}{Zeng, Z.}, \bibinfo{year}{2024}.
\newblock \bibinfo{title}{Analytic federated learning}.
\newblock \bibinfo{journal}{arXiv preprint arXiv:2405.16240} .
%Type = Article
\bibitem[{Sabah et~al.(2024)Sabah, Chen, Yang, Azam, Ahmad and Sarwar}]{sabah2024model}
\bibinfo{author}{Sabah, F.}, \bibinfo{author}{Chen, Y.}, \bibinfo{author}{Yang, Z.}, \bibinfo{author}{Azam, M.}, \bibinfo{author}{Ahmad, N.}, \bibinfo{author}{Sarwar, R.}, \bibinfo{year}{2024}.
\newblock \bibinfo{title}{Model optimization techniques in personalized federated learning: A survey}.
\newblock \bibinfo{journal}{Expert Systems with Applications} \bibinfo{volume}{243}, \bibinfo{pages}{122874}.
%Type = Article
\bibitem[{Gao et~al.(2023)Gao, Wang, Liu, Zhang and Ma}]{gao2023configure}
\bibinfo{author}{Gao, Y.}, \bibinfo{author}{Wang, P.}, \bibinfo{author}{Liu, L.}, \bibinfo{author}{Zhang, C.}, \bibinfo{author}{Ma, H.}, \bibinfo{year}{2023}.
\newblock \bibinfo{title}{Configure your federation: hierarchical attention-enhanced meta-learning network for personalized federated learning}.
\newblock \bibinfo{journal}{ACM Transactions on Intelligent Systems and Technology} \bibinfo{volume}{14}, \bibinfo{pages}{1--24}.
%Type = Article
\bibitem[{Yang et~al.(2023)Yang, Huang, Lin and Cao}]{yang2023personalized}
\bibinfo{author}{Yang, L.}, \bibinfo{author}{Huang, J.}, \bibinfo{author}{Lin, W.}, \bibinfo{author}{Cao, J.}, \bibinfo{year}{2023}.
\newblock \bibinfo{title}{Personalized federated learning on non-iid data via group-based meta-learning}.
\newblock \bibinfo{journal}{ACM Transactions on Knowledge Discovery from Data} \bibinfo{volume}{17}, \bibinfo{pages}{1--20}.
%Type = Article
\bibitem[{T~Dinh et~al.(2020)T~Dinh, Tran and Nguyen}]{t2020personalized}
\bibinfo{author}{T~Dinh, C.}, \bibinfo{author}{Tran, N.}, \bibinfo{author}{Nguyen, J.}, \bibinfo{year}{2020}.
\newblock \bibinfo{title}{Personalized federated learning with moreau envelopes}.
\newblock \bibinfo{journal}{Advances in neural information processing systems} \bibinfo{volume}{33}, \bibinfo{pages}{21394--21405}.
%Type = Article
\bibitem[{Zhuang et~al.(2025)Zhuang, Lin, Yang and Toh}]{zhuang2025analytic}
\bibinfo{author}{Zhuang, H.}, \bibinfo{author}{Lin, Z.}, \bibinfo{author}{Yang, Y.}, \bibinfo{author}{Toh, K.A.}, \bibinfo{year}{2025}.
\newblock \bibinfo{title}{An analytic formulation of convolutional neural network learning for pattern recognition}.
\newblock \bibinfo{journal}{Information Sciences} \bibinfo{volume}{686}, \bibinfo{pages}{121317}.
%Type = Inproceedings
\bibitem[{Toh(2018)}]{toh2018learning}
\bibinfo{author}{Toh, K.A.}, \bibinfo{year}{2018}.
\newblock \bibinfo{title}{Learning from the kernel and the range space}, in: \bibinfo{booktitle}{2018 IEEE/ACIS 17th International Conference on Computer and Information Science (ICIS)}, \bibinfo{organization}{IEEE}. pp. \bibinfo{pages}{1--6}.
%Type = Article
\bibitem[{Ranganathan and Lewandowski(2020)}]{ranganathan2020zorb}
\bibinfo{author}{Ranganathan, V.}, \bibinfo{author}{Lewandowski, A.}, \bibinfo{year}{2020}.
\newblock \bibinfo{title}{Zorb: A derivative-free backpropagation algorithm for neural networks}.
\newblock \bibinfo{journal}{arXiv preprint arXiv:2011.08895} .
%Type = Inproceedings
\bibitem[{Lee et~al.(2023)Lee, Kim and Kim}]{lee2023gpil}
\bibinfo{author}{Lee, G.}, \bibinfo{author}{Kim, N.J.}, \bibinfo{author}{Kim, H.}, \bibinfo{year}{2023}.
\newblock \bibinfo{title}{Gpil: Gradient with pseudoinverse learning for high accuracy fine-tuning}, in: \bibinfo{booktitle}{2023 IEEE 5th International Conference on Artificial Intelligence Circuits and Systems (AICAS)}, \bibinfo{organization}{IEEE}. pp. \bibinfo{pages}{1--5}.
%Type = Article
\bibitem[{{Li Deng}(2012)}]{li_deng_mnist_2012}
\bibinfo{author}{{Li Deng}}, \bibinfo{year}{2012}.
\newblock \bibinfo{title}{The {MNIST} {Database} of {Handwritten} {Digit} {Images} for {Machine} {Learning} {Research} [{Best} of the {Web}]}.
\newblock \bibinfo{journal}{IEEE Signal Processing Magazine} \bibinfo{volume}{29}, \bibinfo{pages}{141--142}.
\newblock \DOIprefix\doi{10.1109/MSP.2012.2211477}.
%Type = Article
\bibitem[{Krizhevsky et~al.(2009)Krizhevsky, Hinton et~al.}]{krizhevsky2009learning}
\bibinfo{author}{Krizhevsky, A.}, \bibinfo{author}{Hinton, G.}, et~al., \bibinfo{year}{2009}.
\newblock \bibinfo{title}{Learning multiple layers of features from tiny images} .
%Type = Article
\bibitem[{LeCun et~al.(1998)LeCun, Bottou, Bengio and Haffner}]{lecun1998gradient}
\bibinfo{author}{LeCun, Y.}, \bibinfo{author}{Bottou, L.}, \bibinfo{author}{Bengio, Y.}, \bibinfo{author}{Haffner, P.}, \bibinfo{year}{1998}.
\newblock \bibinfo{title}{Gradient-based learning applied to document recognition}.
\newblock \bibinfo{journal}{Proceedings of the IEEE} \bibinfo{volume}{86}, \bibinfo{pages}{2278--2324}.
%Type = Article
\bibitem[{Li et~al.(2020)Li, Sahu, Zaheer, Sanjabi, Talwalkar and Smith}]{li2020federated}
\bibinfo{author}{Li, T.}, \bibinfo{author}{Sahu, A.K.}, \bibinfo{author}{Zaheer, M.}, \bibinfo{author}{Sanjabi, M.}, \bibinfo{author}{Talwalkar, A.}, \bibinfo{author}{Smith, V.}, \bibinfo{year}{2020}.
\newblock \bibinfo{title}{Federated optimization in heterogeneous networks}.
\newblock \bibinfo{journal}{Proceedings of Machine learning and systems} \bibinfo{volume}{2}, \bibinfo{pages}{429--450}.
%Type = Article
\bibitem[{Zhang et~al.(2025)Zhang, Chen, Lin, Chen and Zhao}]{zhang2025lcfed}
\bibinfo{author}{Zhang, Y.}, \bibinfo{author}{Chen, H.}, \bibinfo{author}{Lin, Z.}, \bibinfo{author}{Chen, Z.}, \bibinfo{author}{Zhao, J.}, \bibinfo{year}{2025}.
\newblock \bibinfo{title}{Lcfed: An efficient clustered federated learning framework for heterogeneous data}.
\newblock \bibinfo{journal}{arXiv preprint arXiv:2501.01850} .
%Type = Inproceedings
\bibitem[{Chen et~al.(2024)Chen, Zhang, Zhao, Wang and Xu}]{chen2024gradient}
\bibinfo{author}{Chen, H.}, \bibinfo{author}{Zhang, Y.}, \bibinfo{author}{Zhao, J.}, \bibinfo{author}{Wang, X.}, \bibinfo{author}{Xu, Y.}, \bibinfo{year}{2024}.
\newblock \bibinfo{title}{Gradient free personalized federated learning}, in: \bibinfo{booktitle}{Proceedings of the 53rd International Conference on Parallel Processing}, pp. \bibinfo{pages}{971--980}.

\end{thebibliography}

% % Here goes the biography details.

\bio{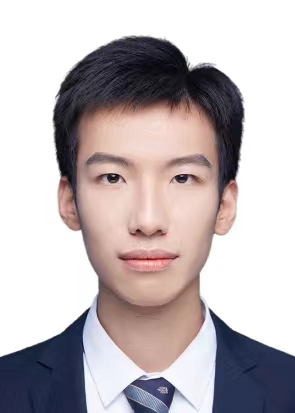}
{Shunxian Gu} received the B.S degree in Science and Technology of Intelligence from Hohai University, Nanjing, China, in 2023. Currently, he is an M.Sc. student at College of Systems Engineering at National University of Defense Technology, Changsha, China. He is also a visiting student at Fudan University, Shanghai, China. His research interests include parallel and high performance computing.
\endbio

\bio{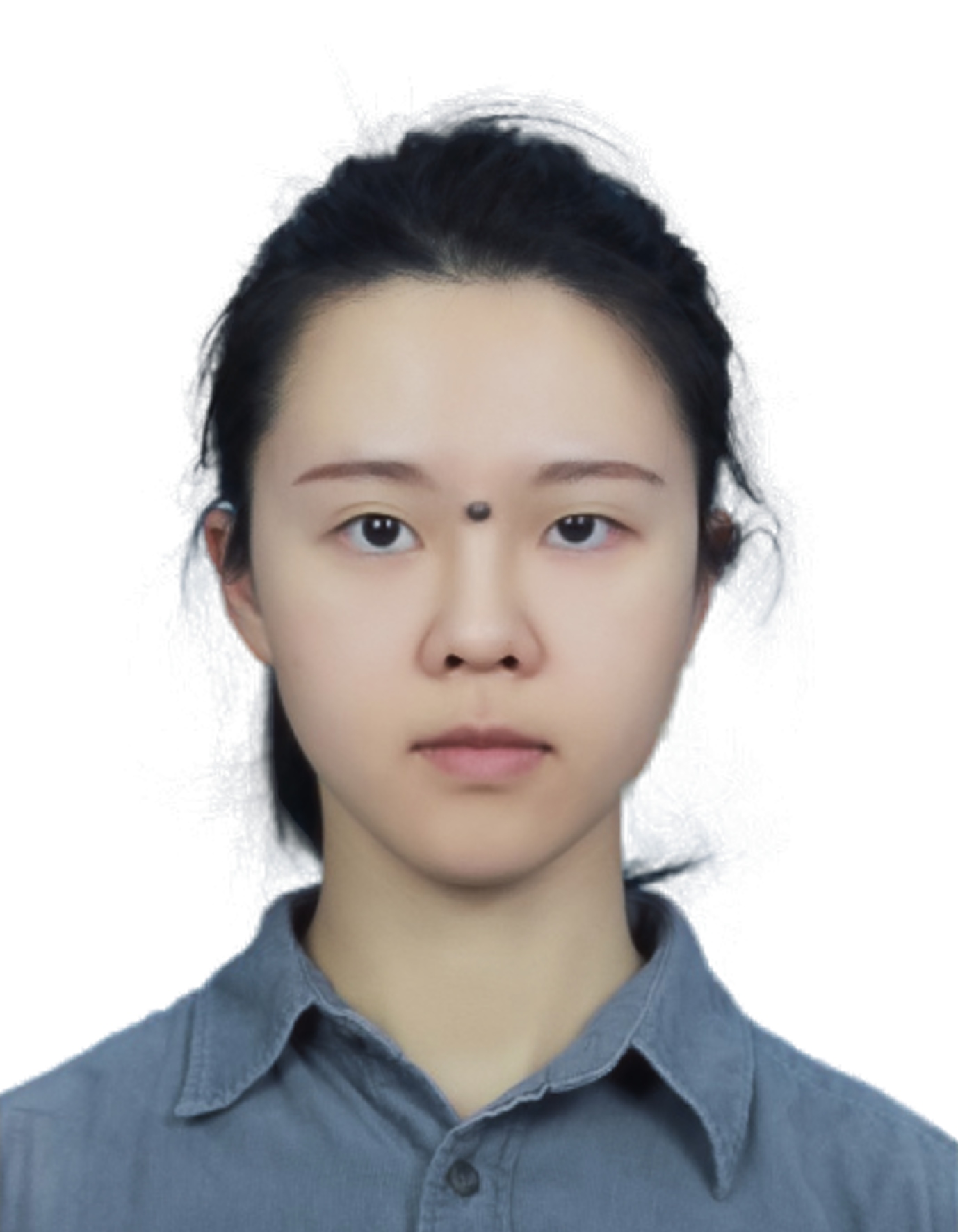}
{Chaoqun You} is an assistant professor at Fudan University. Prior to that, she was a postdoctoral research fellow in Singapore University of Technology and Design (SUTD). She received the B.S. degree in communication engineering and the Ph.D. degree in communication and information system from University of Electronic Science and Technology of China (UESTC) in 2013 and 2020, respectively. She was a visiting student at the University of Toronto from 2015 to 2017. Her current research interests include NTN RAN, network virtualization virtualization, Federated Learning, 5G and 6G.
\endbio

\bio{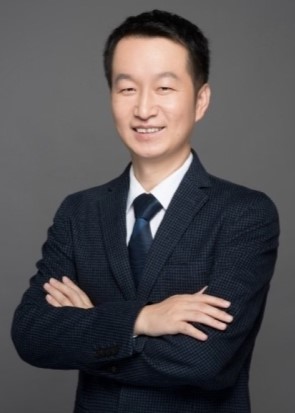}
{Deke Guo} received the B.S. degree in industry engineering from the Beijing University of Aeronautics and Astronautics, Beijing, China, in 2001, and the Ph.D. degree in management science and engineering from the National University of Defense Technology, Changsha, China, in 2008. He is currently a Professor with the College of System Engineering, National University of Defense Technology, and is also with the College of Intelligence and Computing, Tianjin University. His research interests include distributed systems, software-defined networking, data center networking, wireless and mobile systems, and interconnection networks. He is a senior member of the IEEE and a member of the ACM.
\endbio

\bio{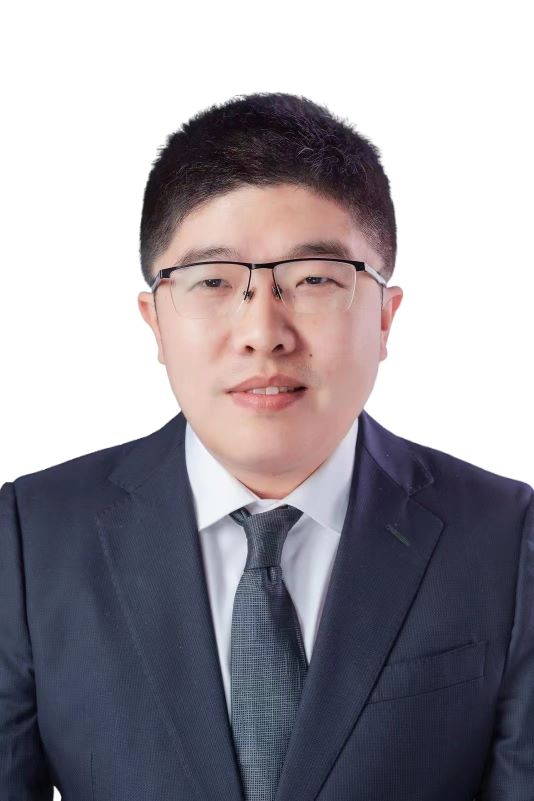}
{Zhihao Qu} received his B.S. and Ph.D. degree in computer science from Nanjing University, Nanjing, China, in 2009, and 2018, respectively. He is currently an associate professor in the School of Computer and Information at Hohai University. His research interests are mainly in the areas of federated learning, edge computing, and distributed machine learning.
\endbio

\bio{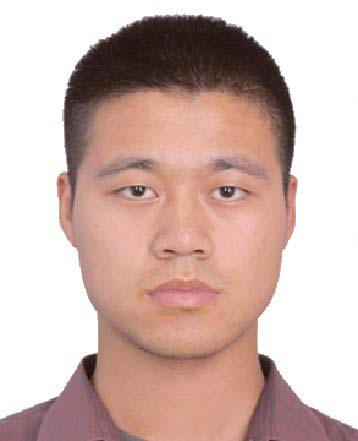}
{Bangbang Ren} received his the B.S., M.S., and Ph.D. degrees from the National University of Defense Technology, China, in 2015, 2017, and 2021, respectively. He was also a Visiting Research Scholar with the University of G\"{o}ttingen, G\"{o}ttingen, Germany, in 2019. His research interests include software-defined network and network optimization.
\endbio

\bio{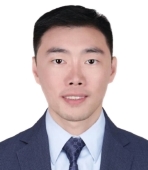}
{Zaipeng Xie} received his B.Eng in Information Engineering from Southeast University, Nanjing, China, in 2002, and his Ph.D. in Electrical Engineering from the University of Wisconsin-Madison, USA, in 2015. He currently holds the position of Associate Professor within the Department of Computer Science and Technology at Hohai University, Nanjing, China. Presently, Dr. Xie's research interests center around distributed machine learning systems and their implications for sustainability.
\endbio

\bio{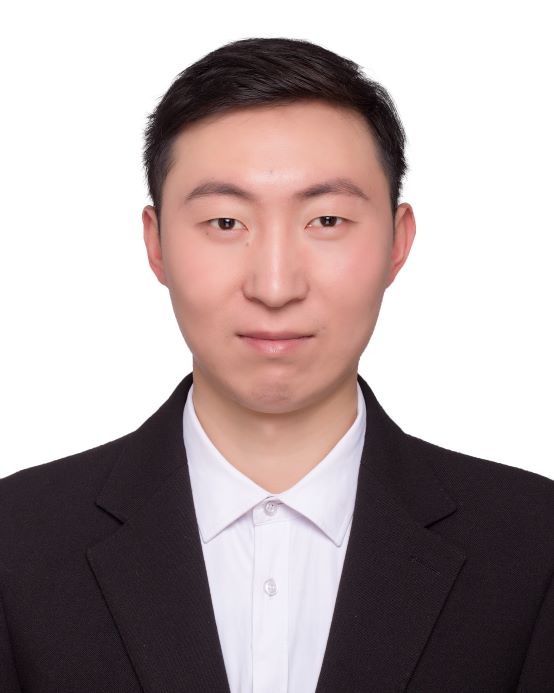}
{Lailong Luo} received his B.S., M.S. and Ph.D degree at the College of Systems Engineering from National University of Defence Technology, Changsha, China, in 2013, 2015, and 2019 respectively. He is currently an associate professor in the School of Systems, National University of Defense Technology, Changsha, China. His research interests include data structure and distributed networking systems. 
\endbio

\end{document}